\documentclass[a4paper]{article}
\usepackage{Odyssey2022}
\usepackage{epsfig,amssymb,amsmath,color,xcolor,enumitem}
\ninept % !!!

\usepackage[english]{babel}
\usepackage[T1]{fontenc}
\renewcommand{\vec}[1]{\boldsymbol{\mathrm{#1}}}
\newcommand{\mtx}[1]{\boldsymbol{\mathrm{#1}}}
\newcommand{\transp}{\ensuremath{^\mathsf{T}}}

\usepackage{appendix}
\usepackage{multirow}

\newcommand\E{\mathbb{E}}

\renewcommand{\vec}{\boldsymbol} % Vector
\newcommand*\dif{\mathop{}\!\mathrm{d}}

\usepackage{algorithmicx} %
\usepackage{algorithm}
\usepackage{algpseudocode}
\usepackage{hyperref}

\usepackage[
backend=biber,
style=ieee,
maxbibnames=3,
maxcitenames=3,
doi=false,isbn=false,url=false,eprint=false
]{biblatex}
\defbibheading{bibliography}[\refname]{}

\renewcommand{\vec}{\boldsymbol} % Vector

\title{Baselines and Protocols for Household Speaker Recognition}%\\ Algorithms, Baselines, Protocols, and Metrics}

\makeatletter
\def\name#1{\gdef\@name{#1\\}}
\makeatother
\name{{\em Alexey Sholokhov$^1$, Xuechen Liu$^2{}^,{}^3$, Md Sahidullah$^3$, Tomi Kinnunen$^2$}}
\address{$^1$Independent researcher \\
$^2$School of Computing, University of Eastern Finland, Finland \\
{$^3$}Universit\'{e} de Lorraine, CNRS, Inria, LORIA, F-54000, Nancy, France \\
\small \tt sholokhovalexey@gmail.com, \{tomi.kinnunen,xuechen.liu\}@uef.fi, md.sahidullah@inria.fr}
\begin{document}
\maketitle

%\tableofcontents

\setlength{\abovedisplayskip}{2pt}
\setlength{\belowdisplayskip}{2pt}
\setlength{\belowcaptionskip}{-10pt}
%\addtolength{\parskip}{-0.5mm}

%
\begin{abstract}
Speaker recognition on household devices, such as smart speakers, features several challenges: (i) robustness across a vast number of heterogeneous domains (households), (ii) short utterances, (iii) possibly absent speaker labels of the enrollment data (passive enrollment), and (iv) presence of unknown persons (guests).
%, and (v) on-device computation. 
While many commercial products exist, there is less published research
% that excels in its absence. 
and no publicly-available evaluation protocols or open-source baselines. Our work serves to bridge this gap by providing %through
%Our main purpose is to %define the scope and to 
%provide 
an accessible evaluation benchmark derived from public resources (VoxCeleb and ASVspoof 2019 data) along with a preliminary pool of %several 
open-source baselines. This %preliminary pool %of baselines 
includes four algorithms for active enrollment (speaker labels available) and one algorithm for passive enrollment.
\end{abstract}

\section{Introduction}
\vspace{-0.15cm}

%\tableofcontents

%\textcolor{red}{IGNORE THIS SECTION FOR THE TIME BEING.}
Speech technology has entered our living rooms. Be it a smart TV or smart home assistant connected to other devices, an increasing number of commercial products now support interaction by voice. In addition to recognizing voice commands correctly (speech recognition), it is important to be aware of the user (speaker identification) for customized experiences such as personalized playlists, recommendations and bookings.

We address such \emph{household speaker recognition} task \cite{Chen2021-Label-Propagation,Tan2021-shared-device-subset} where a small group of users (household members) share a common device. The task is differentiated from generic speaker recognition tasks (addressed in standard benchmark data such as NIST SREs \cite{GREENBERG2020-two-decades} and VoxCeleb \cite{voxceleb1,voxceleb2}) with several desiderata, intended to maximize user experience:

    \begin{footnotesize}
    \begin{itemize}
        \item \textbf{Domain robustness.} The device should be readily-usable in previously unseen domains (households);   
        \item \textbf{User convenience.} User registration (speaker enrollment) should be done as unobtrusively as possible --- ideally without any interaction with the user; %Such \emph{passive enrollment} lacks supervision (speaker labels);
        \item \textbf{Adaptivity.} To maintain high performance through weeks, months, or years, the model(s) require updating (again without user interaction);
        \item \textbf{Short utterances.} The device should be attentive to the user with very short utterances (1-2 second range);
        \item \textbf{Open-set operation.} The device should be attentive only to the members of a given household, i.e. the ability to ignore commands from unknown (out of the household) users;
        \item \textbf{Computational efficiency.} With growing privacy concerns, it might be preferable to carry out \emph{on-device} computations without sending the user's speech data elsewhere.
    \end{itemize}
    \end{footnotesize}
The above translate to computationally light and ideally \emph{unsupervised} (no speaker labels required) approaches that evolve with time, as opposed to single-shot optimization on a given corpus, domain, or challenge. 

In generic speaker recognition setups, the issue of speech variation across different datasets is typically addressed using \emph{domain adaptation} (DA) of an already well-trained recognizer. Practice has shown that it is effective to first train a large speaker embedding network (that serves as a utterance-level speaker feature extractor) \cite{ecapa-tdnn,Snyder_etdnn_2019}, fix it across domains, followed by adaptation of back-end classifier (speaker comparator) to each domain of interest \cite{coral,coralplus,coral_asv2018,kaldi_aplda}.

While DA methods are designed to address \emph{small number of large domains} (many speakers), the household scenario represents the opposite --- \emph{large number of small domains} (households). %While the household size is typically small --- which sets high speaker identification performance \emph{a priori} --- there are potentially millions of different households all with differing acoustics. 
%This sets challenges for the speaker recognition system to operate consistently from household to household. %\textcolor{red}{todo - links/differences/citations with relevant domain adaptation/transfer learning...}
As noted in \cite{Tan2021-shared-device-subset}, systems designed to address \emph{generic} speaker recognition task on large population might be suboptimal on \emph{specific} speaker subsets with homogeneous data. Homogeneous data can indeed be expected due to the shared device with the fixed (or infrequently changed) location. The household scenario is therefore differentiated from speaker recognition \emph{across devices} or \emph{on mobile devices}. In addition, DA methods are typically not designed to handle sequential (online) data arriving one utterance at a time. For these reasons, we consider lightweight back-ends with as few parameters as possible and instead focus on adapting \emph{speaker-specific representations} (here, deep neural embeddings). %As a minor technical novelty, we consider a probabilistic linear discriminant analysis (PLDA) back-end \cite{Brummer2010-two-cov,Sizov2014-unifying} with spherical covariances. 
%As we will show, this back-end is equivalent with the nowadays-popular cosine scoring rule, with the added benefit of...\textcolor{red}{Alexey I need your help to complete the sentence :) }
%As we show, this back-end is equivalent with cosine similarity scoring in the case of a single enrollment utterance, with an added benefit of improved score calibration for multiple enrollment utterances.

Speaker recognition for household scenarios is a relatively recent direction (at least in the published literature) \cite{Chen2021-Label-Propagation,Tan2021-shared-device-subset}. In \cite{Chen2021-Label-Propagation}, enrichment of speaker-specific representation using unlabeled data is achieved through semi-supervised \emph{label propagation} \cite{Zhou2003-local-global}. 
%Another study \cite{Tan2021-shared-device-subset} considers a neural back-end classifier with low-dimensional...\textcolor{red}{anyone please help to write a few more words about that study} 
%Another study \cite{Tan2021-shared-device-subset} considers a neural back-end classifier that learns low-dimensional embeddings optimized for a given speaker subset. 
Another study \cite{Tan2021-shared-device-subset} considers a neural back-end classifier with lower-dimensional space, into which the extracted embeddings are mapped. This space is optimized for a given speaker subset. %\textcolor{blue}{
Even if we draw inspiration from these prior studies, there is an important difference in the assumptions. Specifically,  we argue that in households one must prepare for the possibility of external non-target speakers, or \emph{guests}. Therefore, the label propagation approach in \cite{Chen2021-Label-Propagation} is not directly applicable for potential unknown (out of household) speakers.%}
%Even if we draw inspiration from these prior studies, there are several important differences in the assumptions. \textcolor{red}{(TO BE VERIFIED) First, the label propagation approach in \cite{Chen2021-Label-Propagation} is not directly applicable for potential unknown (out of household) speakers. We argue that in households one must prepare for the possibility of external non-target speakers, or \emph{guests}. Second, ... .}

%\textcolor{red}{
%Things to remember
%\begin{itemize}
 %   \item cite some open-set speaker id studies and NIST SRE04 (?)  unsupervised speaker adaptation works (order of trials matter) for historical context. Also SRE16 and others - domain adaptation 
%\end{itemize}}

We summarize our contributions as follows. First and foremost, given the sparse literature and lacking common methodological standards, % or baselines, established protocols, and metrics, 
we target our study to researchers and practitioners who look for an accessible and reproducible starting point to address household speaker recognition tasks. The methodology sections, therefore, provide a guided taxonomy of methods and suggested terminologies. Importantly, we make our experimental pipeline, methods, and protocols public\footnote{\url{https://github.com/underdogliu/household-speaker-recognition}} both for reproducibility and for promotion of further research on this topic.

Besides advancing \textbf{evaluation infrastructure} as described above, we have a number of algorithmic and experimental design-related contributions too. In terms of \textbf{algorithms}, %\textcolor{blue}{
we modified several existing clustering algorithms including the one from \cite{Diez2020-VBx} by introducing an extra ``background class'' and allowing for partially labeled data. %} 
%\textcolor{red}{Alexey, a few sentences about the modifications/extensions to VB clustering}
Second, we consider a highly-constrained variant of \emph{probabilistic linear discriminant analysis} (PLDA) back-end \cite{Brummer2010-two-cov,Sizov2014-unifying} involving spherical covariances. This results in an equivalent scoring rule with the widely-used cosine similarity in the case of a single enrollment utterance, with an added benefit of improved score calibration for multiple enrollment utterances.
%In terms of \textbf{methods}, we ...\textcolor{red}{new contributions}. 

In terms of \textbf{experimental design}, while we draw inspiration from  \cite{Chen2021-Label-Propagation,Tan2021-shared-device-subset}, we suggest a number of extensions and improvements to the simulation of household speaker recognition tasks. Similar to \cite{Chen2021-Label-Propagation,Tan2021-shared-device-subset} we use the public VoxCeleb data \cite{voxceleb1,voxceleb2} in our experiments. In addition, we leverage another public dataset, ASVspoof 2019 physical access (PA) task \cite{WANG2020101114}. Even if the primary use of this dataset is that of benchmarking spoofing countermeasures, in this work we focus on \emph{bonafide} scenarios (no spoofing attacks). The ASVspoof 2019 PA data consists of carefully controlled simulated environments with variations in room sizes, reverberation time, and talker-to-microphone distances. As opposed to VoxCeleb (crawled from YouTube) where randomly-paired utterances may not be representative of a shared homogenous environment, the ASVspoof 2019 PA data allows a design aligned closer with the assumptions of household scenarios; speakers for a given household are selected from the same simulated environment. The experiments with both simulated environments (ASVspoof 2019) along with `in the wild' style data (VoxCeleb) used in prior studies, therefore, complement one another.

%Last but not least, the purpose of our work is to streamline research in household speaker recognition with the provision of public benchmarking resources for the community. Therefore, we publish all the scripts, household simulation protocols for VoxCeleb and ASVspoof19, and described algorithm implmentations, necessary to reproduce all findings in this paper.

%Differently from prior studies, where speaker model updates are typically constrained to take place with members of the household only, 

%CONTRIBUTIONS
%\begin{itemize}
    %\item Spherical PLDA == cosine scoring (e.g. a short Appendix)
    %\item Prior work uses only household members in adaptation stage; we expand this to include external (unknown non-targets, guest with unknown number of members in the group)
    %\item ASVspoof19 PA, more realistic in terms of data (household share same environment) compared to Amazon 'subset paper' (they used ASV system to define difficulty)
    %\item Open-source/public datasets / emberrers / --> GitHub reproducibility 
%\end{itemize}

\vspace{-0.1cm}
\section{Household Speaker Recognition}
\vspace{-0.1cm}

% Adatation set:
% - includes only known speakers (family members)\\
% - can include unknown speakers (guests)\\
 
% Updates:
% - online --> update speaker model after each utterance/observation \\
% - offline --> wait for more data, update e.g. once per day\\

%\textcolor{red}{We could use this section to give a big picture and set-up the terminology (active/passive enrollment), household members, guests; and maybe cite the prior studies in more detail.}

\subsection{General task set-up and data assumptions}
\vspace{-0.15cm}

Before presenting the methodology details, we set the general stage for the household speaker recognition task. A \textbf{household} is a small set of natural persons (such as family members) whose voice data is processed by a shared device. The task of interest is \emph{open-set speaker identification} (SID) \cite{openset_sid2006} that aims at classifying an unseen utterance $X$ either as one of the $N$ known household members, or an additional $(N+1)$:th `none of the known speakers' class. We refer to such speakers as \textbf{guests}.

At the enrollment (registration) stage, the device collects voice data of the household members to create the speaker-specific model(s). In generic speaker recognition tasks, the speaker labels of enrollment data are typically available and trustworthy. For instance, published corpora and challenge datasets may undergo tedious scrutiny to validate the correctness of speaker labels (and other metadata) before their release. With similar quality control in mind, in the household scenario, the user might be prompted to repeat wake words or read other pre-specified text passages to create a personal profile. As such \textbf{active enrollment} could be time-consuming or otherwise inconvenient, a more attractive alternative is to collect speech data while the device is used and use speaker clustering to infer pseudo speaker labels after a sufficient amount of data has been collected. The user might then be asked for consent for profile creation. Following terminology used in some commercial products, we refer to such procedures as \textbf{passive enrollment}. In this work, we consider both active and passive enrollment strategies. %\textcolor{blue}{To the best of our knowledge, no prior published work has considered the latter scenario.}
% detailed below.

Besides the test and enrollment data, %common to any speaker recognition task, in the household scenario 
we assume an additional set of unlabeled \textbf{adaptation data} to be available for updating speaker-specific models. Such \emph{unsupervised speaker adaptation} was introduced to the NIST speaker recognition evaluation (SRE) series in 2004 \cite{GREENBERG2020-two-decades}. This task is \emph{not} to be confused with techniques such as transfer learning \cite{ts_learning, tslearning2019} or domain adaptation \cite{coral,coralplus,coral_asv2018,kaldi_aplda} that re-train or modify selected parameters of the recognizer itself. In this work, both the speaker embedding extractor and classifier remain fixed whereas speaker-specific models undergo frequent updates. 

Since adaptation data is unlabeled, the decision as to which speaker model (if any) should be updated has to be done by the recognizer automatically. Note that an error-free (\emph{oracle}) SID system leads to an effectively larger amount of enrollment data per user. In any practical case, however, the SID system makes errors, leading to potentially updating incorrect speaker model(s). %Among other factors \emph{the order in which utterances are processed} therefore impacts performance. The ideal speaker adaptation method should be both insensitive to (reasonable) label prediction errors and the order of utterances.

One of the differences between the household speaker recognition from generic speaker recognition evaluation setups is the varying amount of data available for speaker enrollment and model updating. In contrast to, \emph{e.g.}, VoxSRC challenge setup with a single enrollment utterance per trial \cite{voxsrc-2019}, in the household scenario we assume that multiple utterances can be used to construct the speaker model. Further, the number of utterances may vary from speaker to speaker.
%In more precise terms, 
This corresponds to \emph{multi-enrollment} speaker verification \cite{Rajan2014-single-to-multiple, Zeng2021-attmultienroll}.
%Since multiple utterances may be availables ..., it allowes more alternatives for choosing how to represent the speakers and which scoring back-end to use.

\subsection{General procedure}
\vspace{-0.15cm}

Moving on from the data requirements towards methodology, the core task of interest is pairwise comparison of test utterance with hypothesized model(s)\footnote{\emph{Closed-set} speaker identification with $N$ known speakers consists of $N$ pairwise model-test comparisons, and the open-set SID combines this process with additional score thresholding.}. In this study, we use different \emph{deep neural embeddings} \cite{Speaker-rec-deep-learning-review-2021} to represent speech utterances. In particular, utterance $X$ of arbitrary duration is represented using a speaker embedding $\vec{x} \in \mathbb{R}^d$ of $d$ dimensions: $\vec{x} = \texttt{dnn}(X)$. Here $\texttt{dnn}(\cdot)$ is either E-TDNN \cite{Snyder_etdnn_2019}, ECAPA-TDNN \cite{ecapa-tdnn} or a Resnet34 (details in Section \ref{subsec:speaker-embedders}) used for processing both enrollment and test data. %We focus on the classifier (back-end) and keep these embedding extractors fixed.
We consider these embedding extractors fixed and focus on the system back-end.

The underlying computational task is that of creating a \textbf{speaker representation} (speaker model) for each household member using enrollment and speaker adaptation data. We consider two types of speaker representations, based on \emph{sets} and \emph{centroids}. In the former, a speaker is represented by a collection of embeddings extracted from the utterances of that speaker. In the latter, in turn, we maintain only a single averaged embedding (centroid) per speaker. %(average vector) of this set. %\footnote{To be precise, the stored centroids are augmented by counts -- cardinalities of each set.}.
%With $N_\text{e}$ enrollment utterances, the set representation uses $N_\text{e}\cdot d$ parameters and the centroid presentation $d$ parameters to represent a speaker. 
One can always switch from the set representations to centroids, but not in the other direction. We used one of these representations depending on the speaker adaptation algorithm.

In terms of adaptation algorithms, 
we consider several alternatives %approaches, %in Section \ref{sec:algorithms}, 
categorized either as \textbf{online} (sequential) or \textbf{offline} (batch) methods. Whereas the former (Section \ref{sec:online-algorithms}) assumes the unlabeled adaptation data to arrive utterance-by-utterance, the latter (Section \ref{sec:offline-algorithms}) assumes the adaptation data of all speakers to be available at once. While the former might be preferable from a computational point of view, the latter has the potential to benefit from the global structure of adaptation data. A common ingredient to all considered adaptation algorithms is \textbf{scoring function} (Section \ref{sec:models-and-scoring}) that produces a speaker similarity score between $\vec{x}$ and hypothesized speaker model(s).

\begin{algorithm}[!]
\caption{Active enrollment with online updates}
\begin{algorithmic}[]
    \Require{\\
        \begin{itemize}[leftmargin=*]
            \item Labeled enrollment data $\mathcal{X}_\text{enroll}=\{(\vec{x}_i,y_i)\}$ from one household with speakers $y_i \in \{1,2,\dots,K\}$.
            \item Unlabeled adaptation data $\mathcal{X}_\text{a}=\{\vec{x}_t\}$ containing both household members and guests.
            %\item Labeled evaluation data from the household $\mathcal{X}_\text{eval,house}=\{(\vec{x}_k,y_k)\}$, $y_k\in Y$ and guests $\mathcal{X}_\text{eval,guest}=\{(\vec{x}_m,y_m)\}$, $y_m \notin Y$.
        \end{itemize}
    }
    \Ensure{One model $\vec{\theta}_y$ per speaker $y \in \{1,2,\dots,K\}$.}
    %\Ensure{Target scores, Nontarget scores, Guest scores}
    \For{$y \in \{1,2,\dots,K\}$} \Comment{Enroll each speaker}
        \State{$\vec{\theta}_y \leftarrow $ \texttt{ENROLL}$\Big(\{\vec{x}_i \in \mathcal{X}_\text{e} : y_i = y \}\Big)$}
    \EndFor
    \For{$\vec{x} \in \mathcal{X}_\text{a}$} \Comment{Online model update}
        \State{$k^* \leftarrow \arg\max_{k} \texttt{SCORE}(\vec{x},\vec{\theta}_k) $}
        \If{$\texttt{SCORE}(\vec{x},\vec{\theta}_{k^*}) > \tau$}
            \State{$\vec{\theta}_{k^*} \leftarrow \texttt{UPDATE}(\vec{x},\vec{\theta}_{k^*})$}
        \EndIf
    \EndFor 
\end{algorithmic}
\label{alg:active}
\end{algorithm}
\vspace{-0.1cm}
Algorithm \ref{alg:active} outlines a generic template for active enrollment with online updates: after active speaker enrollment, data arrive one by one. The top-scoring model is updated if the speaker similarity score exceeds an update threshold, and none of the other models are updated. 

Note that the presented generic algorithm assumes discarding presumably out-of-set embeddings without any further processing. However, in principle, these embeddings could be utilized for representing the impostor class corresponding to all the possible guest speakers.

\begin{algorithm}[!h]
\caption{Passive enrollment}
\begin{algorithmic}[]
    \Require{\\
        \begin{itemize}[leftmargin=*]
            \item Unlabeled adaptation data $\mathcal{X}_\text{a}=\{\vec{x}_t\}$ containing both household members and guests.
            
            % \item Unlabeled test data $\mathcal{X}_\text{t}=\{\vec{x}_t\}$ containing possibly unknown speakers.
            
            % \item Pre-defined threshold $\tau$
            
            % \item An adaptation model, parameterized by a set of clusters $\vec{\theta} = \{\vec{\theta}_{1},..., \vec{\theta}_{K}\}$, where $K$ is unknown.
        \end{itemize}
    }
    
    \State{$\vec{\theta} = \{\vec{\theta}_{1},..., \vec{\theta}_{K}\}, K \leftarrow \texttt{CLUSTER}(\{\vec{x}_{t} \in \mathcal{X}_\text{a}\}$)} \\ \Comment{Perform clustering}
            
    % \For{$\vec{x}_t \in \mathcal{X}_\text{t}$} \Comment{Score Assignment}
    %     \State{$k^* \leftarrow \arg\max_{k \in Y} \texttt{SCORE}(\vec{x}_t,\vec{\theta}_k) $}
    %     \If{$\texttt{SCORE}(\vec{x}_t,\vec{\theta}_{k^*}) > \tau$}
    %         % \State{$\vec{\theta}_{k^*} \leftarrow \texttt{UPDATE}(\vec{x}_t,\vec{\theta}_{k^*})$}
    %         \State{$y_{k} = k^*$}
    %     \Else 
    %         \State{$y_{k} = -1$}
    %     \EndIf
    % \EndFor 
\end{algorithmic}
\label{alg:passive}
\end{algorithm}
\vspace{-0.1cm}

%\textcolor{blue}{Xuechen: Active enrollment, offline? Maybe one sentence or two.}

% \textcolor{blue}{Xuechen: Meanwhile, for passive enrollment, with the natural absence of enrollment data, we acquire the adaptation data and perform speaker clustering. Details on the clustering algorithm is described in section 6.3.}

% Some explanation for helping the reader to establish relation between passive enrollment and clustering task, with highlighting the differences. 

%Perhaps, diarization can be mentioned, smth like this: "Since the presented problem setup assumes existing of ``outliers'', one can notice similarity with the speaker diarization (which itself is essentially a clustering problem). In speaker diarization, speakers can be seen as ``target'' clusters while non-speech segments can be seen as ``outliers''. While the non-speech segments may exhibit some group structure, this is absolutely irrelevant to the task of identifying speakers. The same is true in the passive enrollment scenario. Given this similarity of the problem setups, it becomes the natural choice to use one of the performance metrics adopted for speaker diarization. These metrics also have an attractive property of being independent of the size of the outlier class."

%`` ''

In the passive enrollment scenario, one needs to evaluate the ability of the speaker recognition system to identify clusters corresponding to speakers in an unlabeled adaptation set as shown in Algorithm~\ref{alg:passive}. \emph{Evaluating} the success of passive enrollment is therefore similar to the evaluation of general clustering algorithms: given a set of data points, the performance of a clustering algorithm is measured by comparing the predicted groups to the ground-truth partition. There are some minor differences, however. In particular, not all clusters are equally important. Specifically, we assume that there are several dominating clusters in the set, representing the classes of interest. These target clusters represent household members. The aim is to correctly identify only these clusters, while the predicted partitions for the rest of the data set can be arbitrary. In other words, if a pair of guest speakers were incorrectly grouped into a single cluster, it should not affect the performance metric.

\vspace{-0.1cm}
\section{Scoring Back-Ends}\label{sec:models-and-scoring}
\vspace{-0.1cm}

In this section, we discuss considerations related to the choice of speaker representation and 
scoring back-ends. Recent challenges including VoxSRC \cite{voxsrc-2019, voxsrc-2020} and VOiCES 2019 \cite{voices} indicate cosine scoring to be a common choice with speaker embedding extractor networks trained with variants of \emph{angular margin softmax losses} such as AM-Softmax and AAM-Softmax \cite{deng2018arcface, chung2020in}. Another popular choice is PLDA which is more efficient in combination with x-vector embeddings \cite{Snyder2018-xvec, BUT-voxsrc2019}.

As for the latter, we consider PLDA model with the full-rank speaker subspace, in the form of a \emph{two-covariance model} \cite{Brummer2010-two-cov}. 
%It is specified by a couple of covariance matrices: $\mtx{B}$ and $\mtx{W}$, representing between- and within-speaker covariances, respectively. We propose a PLDA model with \emph{spherical covariances}, i.e. with $ \mtx{B} = \sigma_\text{B}^2\mtx{I}$ and  $\mtx{W} = \sigma_\text{W}^2\mtx{I}$ where $\sigma_\text{B}^2$ and $\sigma_\text{W}^2$ are between-speaker and within-speaker variances and $\mtx{I}$ denotes an identity matrix. 
It is specified by two covariance matrices, representing between- and within-speaker covariances. We
choose to use \emph{spherical covariances} (scaled identity matrices). Despite being a highly constrained %special case of 
full-covariance PLDA, this low-parameter version may have an edge in the household speaker recognition scenario. %number of advantages. %First, for speaker embedding extractors trained with angular losses, the parameter-free cosine scoring often outperforms a full-covariance PLDA back-end. %
%But %by using results presented in \cite{Burget2011-discriminatively-trained-PLDA,Rohdin-2014}, 
%it can be shown (details omitted due to lack of space) that the log-likelihood ratio score of spherical-covariance PLDA 
%However, it can be shown that spherical covariance PLDA 
%\emph{is equivalent to cosine scoring} (up to affine transform). Also, 
In particular, a model with only two scalar parameters can be presumed to be more robust against domain mismatch compared to full-covariance PLDA which usually requires domain adaptation to the target domain \cite{Aronowitz2014-idvc, coralplus}.
In our experiments, we use embedding averaging for all %the scoring models, %except for this spherical PLDA approach. 
models apart from this spherical covariance PLDA that uses the so-called \emph{by-the-book scoring} \cite{Rajan2014-single-to-multiple} for multiple enrollment utterances.

It can be shown that the spherical covariance PLDA \emph{is equivalent to cosine scoring}\footnote{This could be the reason why such model, to the authors knowledge, was not mentioned in the existing literature.} in the case of a single enrollment utterance. See Appendix \ref{appendix:plda-cos} for the detailed derivation.

\vspace{-0.1cm}
\section{Online Algorithms}\label{sec:online-algorithms}
\vspace{-0.1cm}

%We start from describing two algorithms 
We now detail the algorithm that processes data sequentially, in an online manner. %That is, the algorithm can update speaker models after receiving a new embedding vector.  
%Both algorithms are extremely simple and mainly differ in their speaker representations: centroid based and set based.
%todo: about online
% centroid based and set based representations
%\subsection{Centroid model with exponential smoothing} \label{sec:centroid-method}
In what we refer to as \emph{centroid model with exponential smoothing},
each of the $K$ speakers in a given household is represented by a \emph{centroid} (average vector) of enrollment data. Call these (initial) centroids $\{ \vec{c}_k^{(0)} \}, k=1,\dots,K$. At each time-step $t=1,2,\dots$ we obtain a new utterance represented by speaker embedding $\vec{x}^{(t)}$, which is compared with all the centroids to obtain $K$ similarity scores $s_1^{(t)},\dots,s_K^{(t)}$. %This algorithm is not tied to any specific back-end scoring model; 
Here these scores are computed either using cosine similarity or PLDA \cite{Brummer2010-two-cov}.
%These scores might be computed using cosine scoring, PLDA, or generally any speaker comparator.
%To be more precise, we assume both $\vec{x}^{(t)}$ and each of the speaker-dependent centroids to have length 1, and score computation done with `cosine scoring', i.e. $s_k^{(t)} = \langle \vec{x}^{(t)}, \vec{c}_k^{(t)}\rangle$ where $\langle \cdot \rangle$ denotes dot product.

Let $k^{(t)}_* \equiv \arg\max_{1 \leq k\leq K} \{ s_k^{(t)}\}$ denote the highest scoring speaker and $s^{(t)}_*$ the corresponding score. We update the centroid of the previous time step if and only if  $s^{(t)}_* > \tau$, by     
    \begin{equation}
        \vec{c}_{k^*}^{(t)} = \alpha \vec{x}^{(t)} + (1-\alpha)\vec{c}_{k^*}^{(t-1)},
        \label{eq:exp-update}
    \end{equation}
where $\tau$ and $\alpha$ are control parameters. Update concerns only the top-scoring centroid and %while other centroids remain unchanged. 
\emph{no} update takes place if $s^{(t)}_* \leq \tau$. 
%And we never touch the centroids of the other speakers, updating concerns only the argmax-speaker.

The update rule \eqref{eq:exp-update} corresponds to exponential smoothing \cite{Brown1963-smoothing} %(see also Appendix),
where the updated centroid is a weighted average of the current embedding %$\vec{x}^{(t)}$ 
and the current centroid. Thus, $0 \leq \alpha \leq 1$ serves as \emph{smoothing factor} that controls the rate at which the centroids evolve. Larger values of $\alpha$ correspond to a stronger emphasis on the most recent observation. Note that simple averaging of embeddings fits this update rule too if one relaxes the assumption of the smoothing factor to be a constant and sets it instead as $\alpha_t = 1 / (t + 1)$. %Intuitively, extreme values of $\alpha$ seem to be sub-optimal in terms of recognition performance of the adapted model. 
In summary, the method has two control parameters $\alpha$ and $\tau$. We address their role in the experimental part. For practical purposes, they %are %of both the smoothing factor $\alpha$ and the threshold $\tau$. Both 
are tuned with the aid of a development set. %on the development set. 

One can see that the update rule \eqref{eq:exp-update} does not track the number of performed updates. However, the information about the number of averaged embeddings is required for the PLDA by-the-book scoring. At the same time, the number of updates may not be an adequate measure of the accumulated information since it depends on the value of the smoothing factor (\emph{e.g.} consider $\alpha \approx 1$). We address this issue by proposing a heuristic rule to define the zero-order statistics (counts) in the case of weighted averaging of embeddings. See Appendix \ref{appendix:plda-centroids} for the detailed discussion.

\vspace{-0.1cm}
\section{Offline Algorithms}\label{sec:offline-algorithms}
\vspace{-0.1cm}

We now turn attention to offline algorithms that operate under the assumption that the whole adaptation set is available at once. Recall that in the active enrollment scenario a small labeled enrollment set is augmented by an unlabeled adaptation set. This naturally fits the formulation of \emph{semi-supervised clustering} \cite{Zhu2005-ssl}, where additional knowledge about the true labels of a subset of data points is assumed. This observation led us to extend two clustering algorithms to their semi-supervised versions, detailed in the next two subsections.
%It should be noted that one can trivially get an online version of any of these algorithms by accumulating the data and running it from scratch each time a new speaker embedding was added. Assuming growing computational complexity with the amount of accumulated data, some applications would not allow such costly updates. However, in the household speaker recognition, where interaction with the device are sparse, this strategy may be allowed.
%To summarize this group of algorithms, let us 
%All the following algorithms are semi-supervised extensions of the existing unsupervised clustering algorithms.

\subsection{Semi-supervised k-means clustering}
\vspace{-0.15cm}
We propose straightforward modifications to the widely known k-means algorithm \cite{AKJain1999-clustering} for the household adaptation setup. 
%First, we modified the traditional (unsupervised) k-means algorithm to ignore the data points with known labels while updating the cluster. 
First, we modified the traditional (unsupervised) k-means algorithm to fix the cluster assignments for a subset of embeddings with known speaker labels.
Second, we introduced as additional ``background class'' to allow data points with no cluster assignments, further referred to as outliers. These data points correspond to the guest speakers. %in the context of household speaker recognition. 
%We will further refer to such points as outliers. 
Specifically, similar to the centroid model described above, we use a pre-defined threshold $\tau$ to determine whether a given data point should be assigned to the closest cluster or labeled as an outlier (a guest). This restricts the clusters to have a bounded diameter. Here, the threshold is a control parameter which should estimated from data.

Similar to the centroid model, the cluster centers (centroids) are initialized by computing speaker-specific average embeddings from the enrollment data.

\subsection{Semi-supervised variational Bayesian clustering}
\vspace{-0.15cm}
Our second proposed offline algorithm can be viewed as variational Bayesian inference \cite{Bishop-MachineLearning2006} applied to a Gaussian mixture model \cite{Corduneanu2001-mixture}. 
In fact, it is a slightly modified version of the clustering algorithm described in \cite{Diez2020-VBx}. %for clustering in the embeddings space.
% ref: bishop https://www.microsoft.com/en-us/research/wp-content/uploads/2016/02/bishop-aistats01.pdf
in which the speaker embeddings are assumed to have been generated by the two-covariance PLDA model \cite{Brummer2010-two-cov} discussed in Section \ref{sec:models-and-scoring}. That is, speakers are modeled by Gaussians with shared covariance but speaker-specific means. By assuming that $K$ speakers are present in a given collection of embeddings, this results in a mixture of $K$ Gaussian components. More details are given in the Appendix \ref{appendix:vb}.

We modified the mixture model described in~\cite{Diez2020-VBx} where each component corresponds to a speaker. In particular, we add an extra label to present `background class'. In practice, we add one $(K+1)$th component to the mixture model to represent the marginal distribution of embeddings. This component is assumed to model all of the unknown (guest) speakers in the household scenario. Also, similar to the semi-supervised k-means clustering described above, we fixed the component assignments for a subset of embeddings with known speaker labels. %More details can be found in \cite{Diez2020-VBx}.
%Detailed derivations and the update rules can be found in \cite{Diez2020-VBx}.

This model requires specifying the prior probabilities corresponding to the mixture components $\vec{\pi}=\{\pi_1,\dots,\pi_{K+1}\}$.
To define the prior probability for the `background class', we apply a threshold $\tau$ as in the %similar to those present in the 
other algorithms. In detail, we define this probability as $\pi_{K+1} = \sigma(\tau)$, where $\sigma(\cdot)$ denotes the sigmoid function. Accordingly, the remaining probabilities $\pi_k$ are initialized such that $\sum_k \pi_k = 1$ for $k=1,\dots,K+1$.

\subsection{Label propagation}
\vspace{-0.15cm}

Inspired by recent related work \cite{Chen2021-Label-Propagation}, we also consider a label propagation approach \cite{Zhou2003-labelprop}. Since the original algorithm is not directly applicable to our \emph{open-set} scenario with potentially unknown speakers, we introduced an additional pre-filtering step. First, all the embeddings that are not sufficiently close to the initial speaker models are labeled as outliers and removed. This outlier filtering is again controlled by a pre-defined threshold $\tau$. After filtering, we assume the remaining set to consist of household members only. Thus, the conventional label propagation algorithm is applied to this set.

\vspace{-0.1cm}
\section{Experimental Setup}
\vspace{-0.1cm}

\subsection{Household speaker recognition protocols}
\vspace{-0.15cm}
We have chosen two public datasets for our experiments: ASVspoof 2019 and VoxCeleb. By drawing inspiration from prior studies \cite{Chen2021-Label-Propagation,Tan2021-shared-device-subset}, we design new evaluation protocols for the household speaker recognition task.

\textbf{ASVspoof 2019:} We consider the bonafide audio data of the \emph{physical access} (PA) subset of ASVspoof 2019 representing different room conditions~\cite{WANG2020101114}. The dataset consists of 27 different room conditions with various room sizes, reverberation times, and talker-to-mic distances. In our experiments, we use the evaluation set consisting of 48 speakers (21 male and 27 female). We set the household sizes as 4, 6, 8, and 10, where each household consists of an equal number of male and female speakers. We create the enrollment list, adaptation list, and trial list considering audio files from the same room condition. The speakers are randomly selected. We also have the same number of guest speakers corresponding to a room but chosen randomly from the remaining speakers. We use three utterances for speaker enrollment, whereas the remaining utterances are used for adaptation. We additionally consider the audio data from the guest speakers for adaptation. The adaptation list contains utterances from the household and guest speakers in a random order. The trial lists consist of only same-gender trials as cross-gender non-target trials are relatively easier than the other. %For PLDA adaptation, we also use bonafide audio data available with the dataset for training the spoofing countermeasures.

\textbf{VoxCeleb:} We consider VoxCeleb1 data consisting of 1251 speakers, where audio data was originally collected from YouTube videos~\cite{voxceleb1}. Following the work reported in~\cite{Tan2021-shared-device-subset}, we have chosen four utterances for enrollment, 10 for evaluation, and 13 for adaptation. Each of the utterances was taken from different videos and we have considered speakers with at least 27 video recordings. Similar to %the configuration adopted for 
ASVspoof 2019 PA, we have gender-balanced households. %with an equal number of male and female speakers. 
The households with sizes 4, 6, 8, and 10 were randomly sampled to create 400 simulated households. Each household has the same number of guests as the number of household members. Similar to %the 
ASVspoof 2019, there are no cross-gender trials. We have created two sets of ASV protocols from VoxCeleb1 with disjoint set of speakers. One of them is used as development set whereas the other set is used as test set. 

\begin{table}[]
    \centering
    \begin{tabular}{|c|c|c|}
    \hline
         Dataset & Target & Non-target (Known/Unknown)  \\ \hline
         VoxCeleb-dev & 27874 & 80126 / 108000 \\ \hline
         VoxCeleb-eval & 27706 & 80294 / 108000 \\ \hline
         ASVspoof 2019 & 75600 & 216000 / 291600  \\ \hline
    \end{tabular}
    \caption{Trial statistics of the datasets used for our experiments. (Known/Unknown refers to the non-target trials from the same household speakers/guests.)}
    \label{tab:trialstatistics}
    \vspace{-0.1cm}
\end{table}

To sum up, we designed three different protocols VoxCeleb-dev, VoxCeleb-eval, ASVspoof with trial statistics summarized in Table~\ref{tab:trialstatistics}. The VoxCeleb-dev protocol serves as the development set and the remaining two are used for final evaluation.

\subsection{Speaker embedding extractors and back-end scoring}\label{subsec:speaker-embedders}
\vspace{-0.15cm}
% \textcolor{green}{XUECHEN to describe all the speaker embedding extractors (including which code and training data). For instance, one paragraph per speaker embedding extractor. In general this part can be quite short.}

% In this section, we describe the three types of speaker embeddings, all of which . 

\textbf{X-vector}. This speaker embedding extractor is based on \emph{Extended TDNN} (E-TDNN) \cite{Snyder_etdnn_2019} x-vector. It is trained on \emph{dev} set of VoxCeleb2 \cite{voxceleb2}, with 5994 speakers. It acquires 80-dimensional mel filterbank outputs as acoustic features. Having the same architecture from \cite{Snyder_etdnn_2019}, we introduce two modifications: 1) We replace statistics pooling with attentive statistics pooling \cite{astats_pooling}; 2) For loss function, we substitute the conventional cross-entropy one with additive angular margin softmax \cite{aam_softmax}. 

\textbf{SpeechBrain}. This pre-trained model is recently proposed by SpeechBrain \cite{speechbrain} with pre-trained model available\footnote{\url{https://huggingface.co/speechbrain/spkrec-ecapa-voxceleb}}, based on %the recently-proposed 
\emph{ECAPA-TDNN} \cite{ecapa-tdnn}, another x-vector variant with a hybrid of convolutional and residual building blocks at layers before the pooling layer. It also imports the aforementioned two modifications and the same %type of 
acoustic features. The pre-trained model is trained on \emph{train} set of VoxCeleb1 \cite{voxceleb1} and \emph{dev} set of VoxCeleb2, which results in 7205 speakers.
Both speaker embedding extractors reached promising performance in our pilot experiments
%\footnote{EER is 2.23\% for E-TDNN and 2.19\% for SpeechBrain ECAPA-TDNN on VoxCeleb1 test set \cite{voxceleb1}, scored via PLDA back-end from Kaldi.}.
\footnote{EER is 2.21\% for E-TDNN and 0.90\% for SpeechBrain ECAPA-TDNN on the VoxCeleb1 test set \cite{voxceleb1}, scored via PLDA back-end from Kaldi and cosine similarity, respectively.}.

\textbf{CLOVA}. This pre-trained model is proposed by and described in \cite{clova}. It is based on ResNet34 architecture \cite{resnet34} and trained on \emph{dev} set of VoxCeleb2 \cite{voxceleb2}, with 5994 speakers. The input acoustic feature for training this model is 64-dimensional mel filterbank, with online data augmentation \cite{clova}. This pre-trained model is provided by CLOVA \footnote{\url{https://github.com/clovaai/voxceleb\_trainer}}. Pilot experiments confirm the promising performance of this model\footnote{Reported EER is 1.18\% on the VoxCeleb1 test set \cite{voxceleb1}.}.

Moreover, when extracting the speaker embeddings, before feeding the input audio into the pre-trained models, we perform trimming with a threshold of -20~dB, followed by random chunking with a length of 2~seconds, elaborating the short-duration property of the audios under the household scenario.

% Moreover, when extracting the speaker embeddings, before feeding the input audio into the pre-trained models, we perform trimming with a threshold of -20~dB, followed by random chunking with a length of 2~seconds.

We used different scoring back-ends depending on the type of embedding. For the x-vector embeddings, we used the conventional back-end including linear discriminant analysis, length normalization, and the full-rank PLDA. For other embeddings cosine similarity or the proposed spherical covariance, PLDA were used.

%\textcolor{red}{Add here description of the PLDA back-ends.}

\subsection{Training details}
\vspace{-0.15cm}
%\textcolor{red}{We need a section to include some information about: 1 -- parameter training (PLDA back-end, algorithm parameters, etc, 2 -- algorithms running configurations)}
%\textcolor{red}{Please include algorithms running configurations.}

\textbf{Active enrollment.} We used a subset of speakers from the VoxCeleb1 \emph{dev} set for setting all the tunable parameters.  To this end, we created two trial lists: one for selecting parameters of the algorithms and scoring back-end via grid-search and the other is for training a global linear calibration \cite{Leeuwen2014-calibration}. Naturally, speakers in this development set do not overlap with the speakers used to create the evaluation protocol. We also used the VoxCeleb2 \emph{dev} set with 5994 speakers for training the full covariance PLDA back-end for the x-vector embeddings. Apart from the PLDA model, it includes the whitening transform with linear dimensionality reduction, followed by length-normalization.

\textbf{Passive enrollment}. We use \emph{agglomerative hierarchical clustering} (AHC)~\cite{esl_textbook} with cosine scoring for passive enrollment. AHC is an iterative method that merges the existing clusters if similarities between the clusters are higher than a predefined threshold. As a primary study on this subject, we employ cosine similarity as the similarity measure.

\subsection{Performance evaluation}
\vspace{-0.15cm}

%\textcolor{red}{TOMI will be improving this part (after understanding the protocol details clearly :) }

\textbf{Active enrollment}. We have a common test set composed of three types of trials. A \emph{target} trial consists of matched enrollment and test identities, \emph{known non-target} trial of two unmatched identities from the household, and \emph{unknown non-target} of unmatched enrollment identity with a guest. We report two different kinds of \emph{equal error rates} (EERs,\%). Both use targets as the positive class while the negative class is either one of the two types of non-targets. EERs are computed from scores pooled across all households. 

\textbf{Passive enrollment}. 
%We first perform scoring on labeled test segments against all household members. 
We first perform scoring on labeled test segments against all detected clusters which are supposed to represent individual speakers.
Similar to adaptation data, the test data includes both household members and guest speakers. Each test segment is labeled with the closest cluster if the corresponding score is above a given threshold. Otherwise, the test segment is labeled as belonging to an `unknown' speaker, that is, not a household member.
%We pool all the scores across all households and employ Jaccard error rate (JER) \cite{jer_original2019} with varying threshold values on score filtering. 
Finally, we pool all the predictions across the households and employ micro-averaged \emph{Jaccard error rate} (JER)~\cite{jer_original2019} designed to measure the quality of clustering results. The numerical values of JER range from 0\% to 100\%. Lower values of JER indicate better clustering. We adopt JER as we are only interested in finding the clusters of household speakers.
For finding the clustering threshold, we first perform a grid search within a certain range and directly perform the evaluation; we then find the optimal value on the subset from VoxCeleb-dev (same as the aforementioned one for the active enrollment) and evaluate with the found threshold. Thresholds are individually optimized for the type of embeddings.
% \textcolor{blue}{Xuechen: add some description on what is JER}.

\vspace{-0.1cm}
\section{Results for Active Enrollment}
\vspace{-0.1cm}

\subsection{Online adaptation with the centroid model}
\vspace{-0.15cm}

We begin by examining the basic properties of the online algorithm described in Section~\ref{sec:online-algorithms}. Specifically, we consider a special case of the update rule \eqref{eq:exp-update} that corresponds to simple averaging --- summing embeddings with equal weights. We consider three different scoring methods, all of which are equivalent to cosine scoring in the case of a single enrollment utterance: cosine scoring with embedding averaging (CSEA), cosine scoring with score averaging (CSSA), and the proposed spherical covariance PLDA (sph PLDA).

\begin{figure}[!h]
\centering
\includegraphics[scale=0.24]{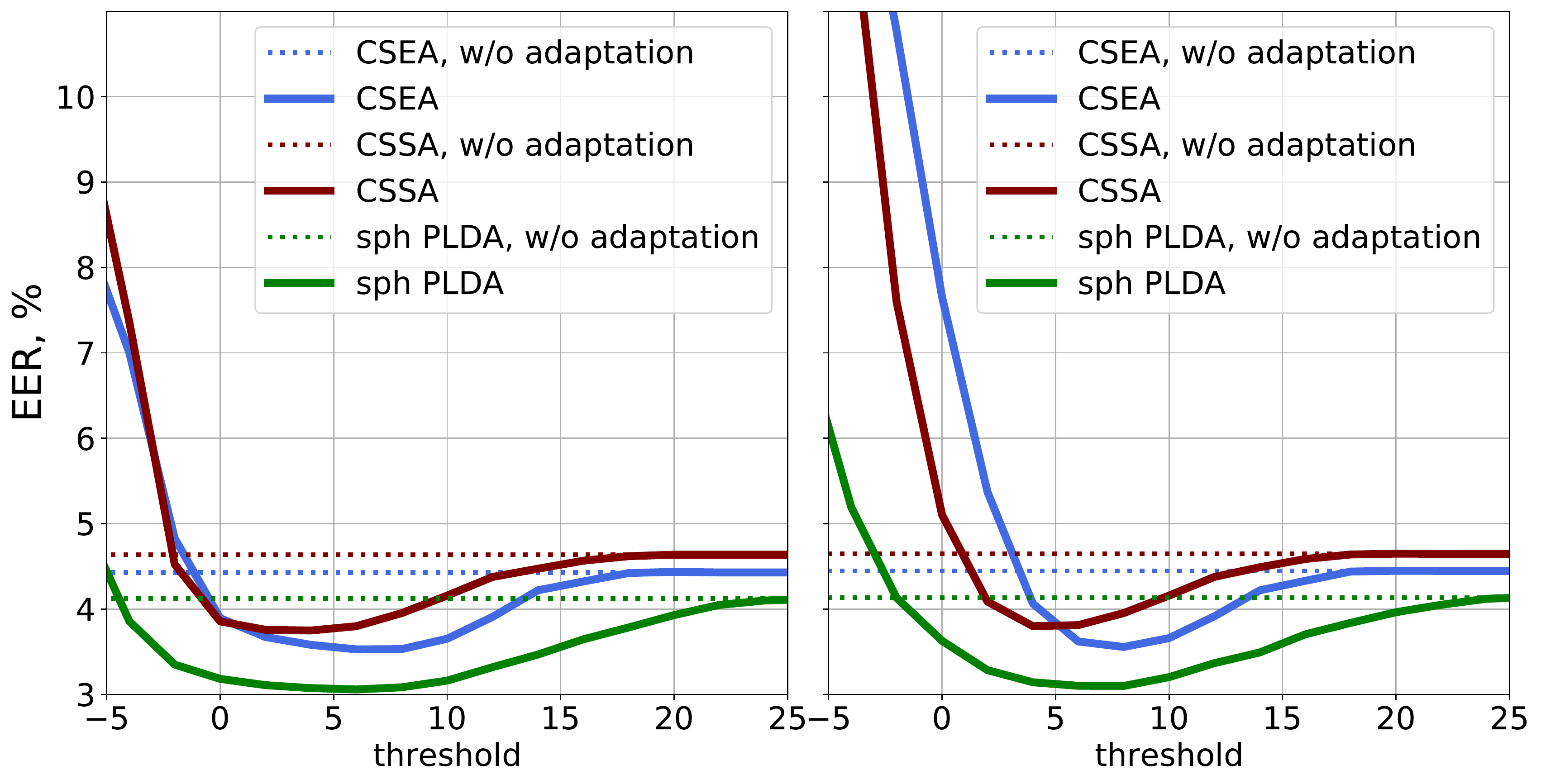}
\caption{Comparison of scoring back-ends on the ASVspoof protocol. All the scores were calibrated by linear logistic regression. The presented results were obtained for the SpeechBrain embeddings. Left: EERs with known non-targets. Right: EERs with unknown non-targets.}
\vspace{-0.1cm}
\label{fig:asvspoof_speechbrain_centroid_scoring_comparison}
\end{figure}

Figure \ref{fig:asvspoof_speechbrain_centroid_scoring_comparison} displays the two different kinds of EERs on ASVspoof data as a function of the model update threshold, $\tau$ (we observed similar trends on VoxCeleb). As can be seen, with a high threshold the metrics approach the case of \emph{no adaptation}. Indeed, if $\tau$ is set too high, no embeddings will be accepted for updating speaker models. On the opposite end, too low $\tau$ may lead to speaker models updated by embeddings of wrong speakers. There is (scoring method dependent) `sweet spot' that leads to substantial improvement over no adaptation. In terms of the scoring approaches, the proposed spherical covariance PLDA %with by-the-book scoring 
achieves lower EERs than either of the cosine scoring variants on the full range of thresholds.
%evaluation protocol.

% \begin{figure}[!h]
% \centering
% \includegraphics[scale=0.24]{fig/proto_v5b_clova_2sec_cos_emb_avg_centroid_alpha.pdf}
% \caption{ASVspoof. Centroid. Alphas}
% \label{fig:proto_v5b_clova_2sec_cos_emb_avg_centroid_alpha}
% \end{figure}

% \begin{figure}[!h]
% \centering
% \includegraphics[scale=0.24]{fig/proto_v5b_speechbrain_trimmed_2sec_cos_emb_avg_centroid_alpha.pdf}
% \caption{ASVspoof. Centroid. Alphas}
% \label{fig:proto_v5b_speechbrain_trimmed_2sec_cos_emb_avg_centroid_alpha}
% \end{figure}

% \begin{figure}[!h]
% \centering
% \includegraphics[scale=0.24]{fig/proto_vox1b_v2_clova_2sec_cos_emb_avg_centroid_alpha.pdf}
% \caption{Voxceleb. Centroid. Alphas}
% \label{fig:proto_vox1b_v2_clova_2sec_cos_emb_avg_centroid_alpha}
% \end{figure}

% \begin{figure}[!h]
% \centering
% \includegraphics[scale=0.24]{fig/proto_v5b_clova_centroid_scoring_comparison.pdf}
% \caption{ASVspoof. Centroid. Scoring comparison}
% \label{fig:asvspoof_clova_centroid_scoring_comparison}
% \end{figure}

% \begin{figure}[!h]
% \centering
% \includegraphics[scale=0.24]{fig/proto_vox1b_v2_clova_centroid_scoring_comparison.pdf}
% \caption{ASVspoof. Centroid. Scoring comparison}
% \label{fig:vox_clova_centroid_scoring_comparison}
% \end{figure}

\begin{table*}[t]
\centering
\begin{tabular}{|l|l|c|c|c|c|c|lll} 
\cline{1-7}
\multirow{2}{*}{Algorithm}   & \multicolumn{1}{c|}{\multirow{2}{*}{Scoring}} & \multicolumn{3}{c|}{ASVSpoof}             & \multicolumn{2}{c|}{VoxCeleb} & \multicolumn{1}{c}{} & \multicolumn{1}{c}{} &   \\ 
\cline{3-7}
                               & \multicolumn{1}{c|}{}                         & CLOVA       & SpeechBrain & x-vector      & CLOVA       & x-vector        &                      &                      &   \\ 
\cline{1-7}
No adaptation                  & CSEA                                          & 3.75 / 3.82 & 4.42 / 4.44 & --             & 1.87 / 1.74 & --               &                      &                      &   \\
No adaptation                  & PLDA                                          & 3.73 / 3.83 & 4.12 / 4.13 & 9.55 / 9.64   & 1.74 / 1.74 & 3.85 / 3.87     &                      &                      &   \\ 
\cline{1-7}
Centroid, $\alpha=1 / (t+1)$   & CSEA                                          & 3.03 / 4.06 & 3.54 / 3.60 & --             & 1.39 / 1.40 & --               &                      &                      &   \\
Centroid, $\alpha=1 / (t+1)$   & PLDA                                          & 3.04 / 3.52 & 3.06 / 3.08 & 11.88 / 17.57 & 1.38 / 1.38 & 3.04 / 3.32     &                      &                      &   \\
Centroid, $\alpha=0.1$         & CSEA                                          & 3.76 / 5.11 & 3.53 / 3.57 & --             & 1.51 / 1.52 & --               &                      &                      &   \\
Centroid, $\alpha=0.1$         & PLDA                                          & 3.41 / 3.98 & 3.10 / 3.15 & 20.31 / 26.94 & 1.36 / 1.35 & 3.05 / 3.38     &                      &                      &   \\ 
\cline{1-7}
k-means                        & CSEA                                          & 3.07 / 4.17 & 3.50 / 3.51 & --             & 1.46 / 1.46 & --               &                      &                      &   \\
k-means                        & PLDA                                          & 3.34 / 5.59 & 3.10 / 3.14 & 12.52 / 18.74 & 1.32 / 1.41 & 3.02 / 3.19     &                      &                      &   \\
VB                             & CSEA                                          & 3.21 / 4.79 & 3.59 / 3.61 & --             & 1.48 / 1.61 & --               &                      &                      &   \\
VB                             & PLDA                                          & 3.31 / 4.96 & 3.14 / 3.17 & 13.48 / 19.31 & 1.34 / 1.45 & 3.21 / 4.03     &                      &                      &   \\
LP                             & CSEA                                          & 3.00 / 3.65 & 3.52 / 3.58 & --             & 1.48 / 1.51 & --               &                      &                      &   \\
\cline{1-7}
Oracle (error-free adaptation) & CSEA                                          & 2.84 / 2.91 & 3.29 / 3.29 & --             & 1.25 / 1.13 & --               &                      &                      &   \\
Oracle (error-free adaptation) & PLDA                                          & 2.91 / 3.01 & 2.79 / 2.77 & 7.94 / 8.03   & 1.25 / 1.20 & 2.69 / 2.74     &                      &                      &   \\ 
\cline{1-7}
\end{tabular}
\vspace{-0.1cm}
\caption{EERs (\%) with known / unknown non-targets for different embedding extractors and algorithms evaluated on the ASVspoof and the VoxCeleb protocols. From top to bottom, the first pair of rows represent the case where the adaptation set is empty. The next block shows results for the online centroid model with two different options to define the smoothing factor. The third block demonstrates different offline algorithms described in section \ref{sec:offline-algorithms}, where VB stands for variational Bayesian clustering and LP is the label propagation. Finally, the last two rows show the Oracle adaptation corresponds to including the adaptation set to the enrollment set. Two scoring back-ends were used: cosine scoring with embedding averaging (CSEA) and PLDA. Since the training set for the SpeechBrain embedding extractor includes the Voxceleb1 data, we did not use it for the second protocol.}
\vspace{-0.1cm}
\label{tab:eer}
\end{table*}
%\vspace{-0.1cm}

% shows the Oracle adaptation corresponds to including the adaptation set to the enrollment set. The next block

\subsection{Comparison of online and offline methods}
\vspace{-0.15cm}

We now compare all the adaptation algorithms --- online and offline --- for different speaker embeddings extractors. The parameters of each algorithm were tuned on the development set (to give the lowest EERs) and kept fixed at the evaluation step. Besides the algorithms described above, we consider two other useful points of reference. The first one, \textbf{no adaptation}, %is
%As a baseline, we measured the EERs for the case if no adaptation set is available, 
which corresponds to a conventional speaker verification evaluation setup without additional adaptation data. The second \mbox{one, \textbf{oracle}}, %Also, we measured the performance of the oracle adaptation algorithm by 
including the adaptation data as part of the enrollment set; this corresponds to an error-free adaptation algorithm that predicts the speaker labels correctly. %of the adaptation set correctly.

%which is able to perfectly predict the ground-truth labels for the adaptation set.

Table \ref{tab:eer} displays the results %of evaluation on two different evaluation protocols: 
on both the ASVspoof and VoxCeleb evaluation protocols. %The results are presented for two different scoring methods, independent of the adaptation algorithms (apart from the centroid and k-means that require to specify a similarity measure to be used). 
The results are presented for two different scoring methods. In general, scoring is independent of the adaptation algorithm, however, in some cases (\emph{e.g.} k-means) the algorithm also uses a similarity function as a part of its decision making process. While that function may be different from the one used for scoring verification trials, we did not consider such options in our experiments.
%the scoring function is a part of the algorithm

First, one can see that adaptation indeed helps to reduce the EER in all cases, apart from x-vector embeddings on the ASVspoof data. This can be explained by the sensitivity of the full-covariance PLDA model to domain mismatch as we did not apply any domain adaptation\footnote{Not to be confused with speaker adaptation studied in this work.} techniques such as \cite{Aronowitz2014-idvc, coralplus}. This is reinforced by noting that x-vector embeddings also benefit from speaker adaptation when evaluated with the matched (VoxCeleb) domain.

Note that the metrics reported in Table \ref{tab:eer} may substantially deviate from the values estimated using in-domain development data. Due to domain mismatch, the threshold tuned on the development set may be suboptimal in other domains. %\textcolor{blue}{
Nonetheless, with motivations noted in Section 1, the authors have purposefully refrained from extensive domain-specific scoring method optimization --- we wanted to expose the methods to handle the unknown. %}.
%It should be noted that the Tables does not show the lowest EERs that could be achieved by the algorithms. Due to domain mismatch the threshold tuned on the development set may be suboptimal for other domains.

Another important and perhaps surprising observation from Table \ref{tab:eer} is that the online algorithm achieved comparable (or sometimes even better) performance to offline algorithms. %\textcolor{blue}{
This finding might be explained by the relatively large amount of data available for adaptation. We hypothesize that such an amount is sufficient for saturation in performance
while lower amounts of adaptation data might benefit from offline methods. %could be beneficial.%}
%Due to the lack of space we leave the question on the effects of varying the adaptation set size for future work.
%Combined with simplicity and computationally cheap updates this makes the centroid model an attractive solution for household speaker recognition.
%Competitive performance together with computationally cheap updates can be sees as main advantages of the online updates.
% desirable property, computationally cheap updates

Finally, one can see that all the offline algorithms have a comparable performance by achieving up to 20\% relative reduction in EER with known non-targets and up to 15\% relative reduction in EER with unknown non-targets. With this observation, a possible choice could be using the arguably simplest algorithm -- semi-supervised k-means with thresholding.

\begin{figure}[!h]
\centering
\includegraphics[scale=0.22]{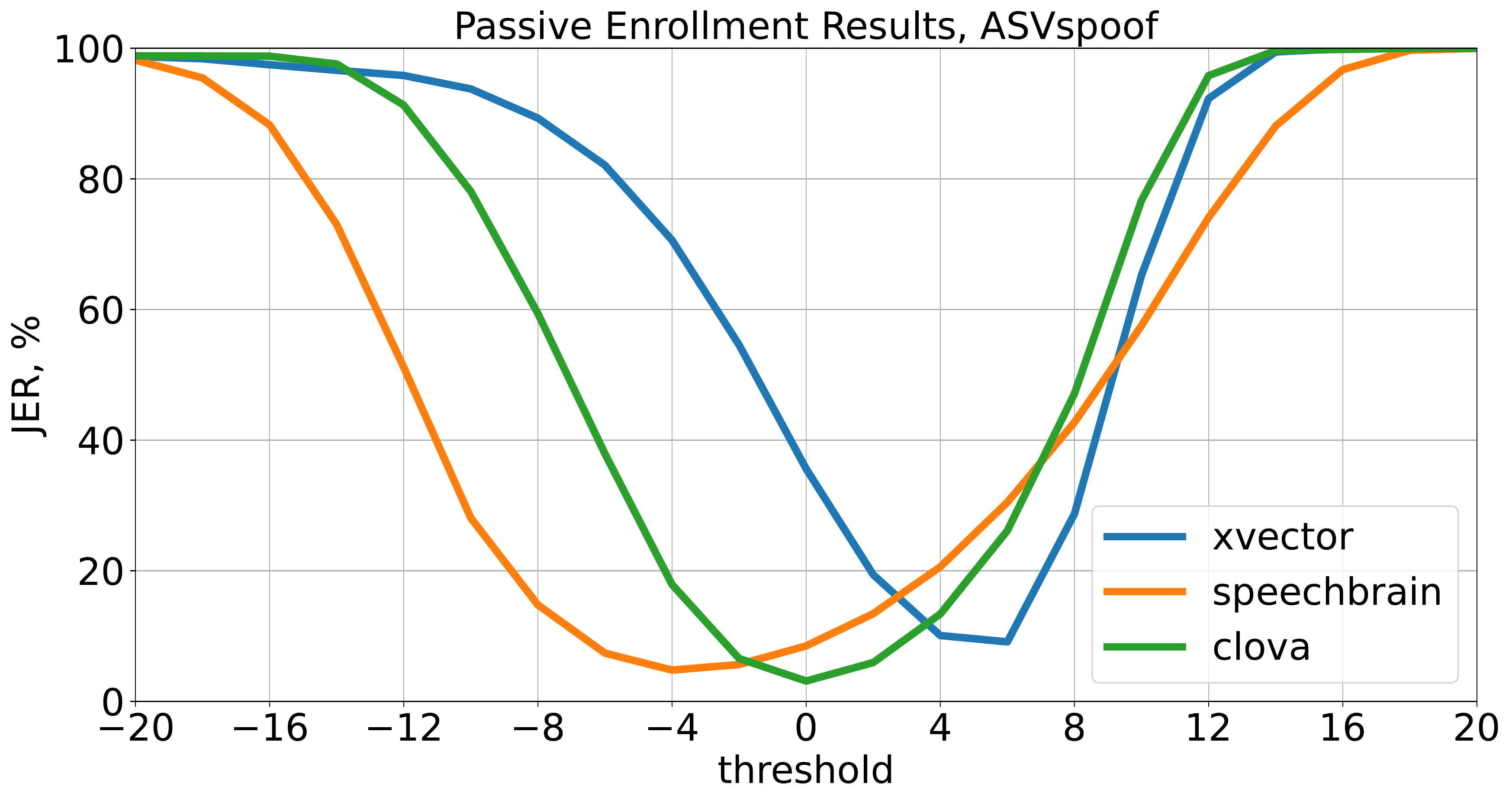}
\vspace{-0.1cm}
\caption{Results (in terms of JER, \%) for passive enrollment with three types of speaker embeddings.}
\label{fig:passive_enrollment}
%\vspace{-0.1cm}
\end{figure}

% 8.86	1.5	4.55

\begin{table}[]
    \centering
    \begin{tabular}{|c|c|c|}
    \hline
         Embedding & ASVspoof & VoxCeleb  \\ \hline
         CLOVA & 27.42 & 4.55 \\ \hline
         SpeechBrain & 7.41 & - \\ \hline
         x-vector & 70.64 & 8.86 \\ \hline
    \end{tabular}
    \vspace{-0.1cm}
    \caption{JERs (\%) of different embeddings on the two evaluation sets. The threshold value is optimized individually. Since the training set for the SpeechBrain embedding extractor includes the VoxCeleb1 data, we do not evaluate it here.}
    \label{tab:passive_enrollment}
    \vspace{-0.1cm}
\end{table}

\vspace{-0.1cm}
\section{Results for Passive Enrollment}
\vspace{-0.1cm}
The results of passive enrollment with varying thresholds are illustrated in Fig.~\ref{fig:passive_enrollment} for the ASVspoof dataset for all three types of speaker embeddings. SpeechBrain yields lower JER than the other two embeddings for a wide range of thresholds. Performance becomes worse till 100\% JER is beyond a certain range of threshold values applied on the similarity measures. For a certain lower value of the threshold, the number of predicted clusters is less, hence resulting in a high false alarm. Conversely, for a certain higher value of the threshold, the number of predicted clusters is very high and that leads to a high miss rate.

%Results of passive enrollment are illustrated in Fig.~\ref{fig:passive_enrollment} for ASVspoof dataset for all the three types of speaker embeddings. For more negative values of threshold, SpeechBrain yields better performance than the other two embeddings. They also holds lowest JER if optimal threshold value can be fetched for each type of embeddings separately. While clova stands out when the threshold value is around zero, for threshold values between 4 and 8, xvector returns lower JER, which indicates its efficiency for systems less strict on false rejection. As the optimal region varies with respect to the three embeddings, future work under this scenario may start from applying fusion or ensemble methods on them.

In our final experiment, we use the VoxCeleb \emph{dev} data to set the thresholds (one per each embedding type). %We separately optimize the thresholds for different embeddings 
These thresholds then remain fixed on the \emph{evaluation} part of ASVspoof and VoxCeleb data. The results in Table~\ref{tab:passive_enrollment} indicate that SpeechBrain performs best on ASVspoof with a JER of 7.41\%, while the performance is worst for the x-vector embedding extractor. This might be due to dataset mismatch leading to a threshold that fails to generalize from VoxCeleb on ASVspoof. Concerning the relative performance of CLOVA and x-vector, the former outperforms the latter by demonstrating better generalization ability on both datasets. Nevertheless, the best performances via grid search over different clustering thresholds (Fig.~\ref{fig:passive_enrollment}) indicate scope for improvement in passive enrollment.

%Table~\ref{tab:passive_enrollment} shows the results returned by the threshold value learned from the VoxCeleb dev subset. SpeechBrain embeddings demonstrate their better performance than the other two, without creating larger performance mismatch between VoxCeleb and ASVspoof than active enrollment counterparts. However, this is not the case for CLOVA and x-vector, where the JERs returned are not as good either. This indicates that SpeechBrain not only is able to provide better performance with optimal threshold via grid search, but also utilizes better generalization ability, under the setup of passive enrollment.

All in all, the above findings are observations that serve as a baseline and future work on passive enrollment may start from here.

\vspace{-0.1cm}
\section{Conclusion}
\vspace{-0.1cm}

% Our contributions, results

The aim of this work has been the creation of 
%In this work we attempted to create 
an accessible and reproducible entry point to address household speaker recognition tasks. Besides %providing %involving a preliminary pool of baseline methods 
% First, we presented 
detailing the task and assumptions, %and relation to other problem formulations, 
we designed a public 
% Also, we introduced a public 
evaluation benchmark for both active and passive enrollment; and online and offline algorithms.
% for updating speaker representations. %While this benchmark allows  
% We evaluated several baselines including online (sequential) and offline algorithms for adapting speaker models.
From the various methods considered, the simple online update rule based on the weighted average of embeddings achieved competitive performance to offline algorithms. The simplicity and computationally cheap updates make it a viable solution for household speaker recognition.% scenarios.
%While we hope our work to inspire further solutions to the 
%We have a number of future plans

%...
Our planned future work includes (1) improving the baselines to improve domain robustness. For instance, it would be interesting to generalize existing domain adaptation methods to handle online data. (2) We also plan to analyze the online methods in terms of their sensitivity to the \emph{order} of data.
(3) Despite the promising preliminary results, it is also important to analyze the impact of the critical control parameters on the convergence properties of the algorithms. An important difference with generic speaker recognition tasks is that the model(s) are allowed to evolve with time which complicates analysis. Nonetheless, it is preferable to find principled alternatives over grid search. 
(4) Finally, we plan to include more methods, especially to the passive enrollment scenarios which we only slightly touched upon. We may also need new performance metrics for a more aligned comparison of active and passive scenarios.

%

%\iffalse

\section{Appendix}
%\appendix

\subsection{PLDA with spherical covariance $\equiv$ cosine scoring} \label{appendix:plda-cos}

Let $\vec{x}_{i,j} \in \mathbb{R}^d$ denote $j$th speaker embedding (observation) of speaker $i$. The \emph{two-covariance} PLDA \cite{Brummer2010-two-cov} assumes these observations to have been generated by a linear Gaussian model of the form
    \begin{equation} \nonumber
        \vec{x}_{i,j} = \vec{\mu} + \vec{y}_i + \vec{\varepsilon}_{i,j},
    \end{equation}
where $\vec{\mu}$ is global mean of the embedding space, $\vec{y}_i$ is latent speaker identity variable with prior distribution $\vec{y}_i \sim \mathcal{N}(\vec{0},\mtx{B})$ and $\vec{\varepsilon}_{i,j}$ a residual with prior $\vec{\varepsilon}_{i,j} \sim \mathcal{N}(\vec{0},\mtx{W})$. The model is equivalently specified by the following distributions:
    \begin{equation}\label{eq:plda-model}
        \begin{aligned}
            p(\vec{y}_{i}) & = \mathcal{N}(\vec{y}_{i}|\vec{0},\mtx{B}) \\ 
            p(\vec{x}_{i,j}|\vec{y}_{i}) & = \mathcal{N}(\vec{x}_{i,j}|\vec{\mu} + \vec{y}_i, \mtx{W}).
        \end{aligned}
    \end{equation}
The two matrices $\mtx{B}$ and $\mtx{W}$ model between- and within-speaker covariances, respectively. By imposing subspace structure on these matrices (rank $<d$) one may obtain other flavors of PLDA \cite{Sizov2014-unifying} as special cases. Here, we consider the highly-constrained case of \emph{spherical covariances},
    \begin{equation} %\nonumber
        \begin{aligned} \label{eq:spherical-assumption}
            \mtx{B} & = \sigma_\text{B}^2\mtx{I}\\
            \mtx{W} & = \sigma_\text{W}^2\mtx{I},
        \end{aligned}
    \end{equation}
where $\sigma_\text{B}^2$ and $\sigma_\text{W}^2$ are between-speaker and within-speaker variances and $\mtx{I}$ denotes an identity matrix.

For speaker comparison, the PLDA model \eqref{eq:plda-model} is used for computing log-likelihood ratio (LLR) score of same-speaker vs. different-speaker hypotheses for a pair of enrollment $(\vec{x}_\text{e})$ and test $(\vec{x}_\text{t})$ embeddings. For the null (same-speaker) hypothesis one assumes $\vec{x}_\text{e}$ and $\vec{x}_\text{y}$ to share common latent identity variable while for the alternative (different-speaker) hypothesis these latent factors are assumed to be different. While the interested reader may refer to details in 
\cite{Burget2011-discriminatively-trained-PLDA,Rohdin-2014}, the resulting LLR score can be written in closed form as
%The likelihood of same-speaker hypothesis is 
	%\begin{equation}\label{eq:target-likelihood}\nonumber
     %   	 L_\text{same}=\mathcal{N}\left( 
      %  	\left[
       % \begin{array}{c}
    %    \vec{x}_\text{e}  \\
    %    \vec{x}_\text{t} 
    %    \end{array}
    %    \right] 	
    %    | 
    %    	\left[
    %    \begin{array}{c}
    %    \vec{\mu}  \\
    %    \vec{\mu} 
    %    \end{array}
    %    \right] ,
    %    	\left[
    %    \begin{array}{cc}
    %    \mtx{B}+\mtx{W} & \mtx{B} \\
    %    \mtx{B} & \mtx{B}+\mtx{W}
    %    \end{array}
    %    \right]
    %    \right),
    %\end{equation}   
%while the alternative (different speaker) likelihood is%
	%\begin{equation}\label{eq:nontarget-likelihood}\nonumber
     %   L_\text{dif} = \mathcal{N}\left( 
      %  	\left[
       % \begin{array}{c}
        %\vec{x}_\text{e}  \\
        %\vec{x}_\text{t} 
        %\end{array}
        %\right] 	
        %| 
        %	\left[
        %\begin{array}{c}
        %\vec{\mu}  \\
        %\vec{\mu} 
        %\end{array}
        %\right] ,
        	%\left[
        %\begin{array}{cc}
        %\mtx{B}+\mtx{W} & \mtx{0} \\
        %\mtx{0} & \mtx{B}+\mtx{W}
        %\end{array}
        %\right]
        %\right).
    %\end{equation}   
%Here, $\mtx{B}+\mtx{W}$ is the \emph{total covariance matrix}. 
%The log-likelihood ratio (LLR), $\ell=\log(L_\text{same}) - \log(L_\text{dif})$, is then used as the speaker similarity score. By expanding the log of the above Gaussian densities and simplifying expressions, the LLR score can be written as \cite{Rohdin-2014}
	\begin{equation} \label{eq:plda-LLR-score-expanded}\nonumber
    	\ell = \vec{x}_\text{e}\transp\mtx{P}\vec{x}_\text{t} + \vec{x}_\text{t}\transp\mtx{P}\vec{x}_\text{e} + \vec{x}_\text{e}\transp\mtx{Q}\vec{x}_\text{e} + \vec{x}_\text{t}\transp\mtx{Q}\vec{x}_\text{t} + \vec{c}\transp (\vec{x}_\text{t} + \vec{x}_\text{e}) + k,
    \end{equation}
where the matrices $\mtx{P}$, $\mtx{Q}$, the vector $\vec{c}$ and scalar $k$ depend only on the PLDA model parameters $(\vec{\mu}, \mtx{B}, \mtx{W})$. With the spherical covariance assumption \eqref{eq:spherical-assumption}, $\mtx{P}$ and $\mtx{Q}$ turn out to be scaled identity matrices too, say $\mtx{P}=\alpha\mtx{I}$, $\mtx{Q}=\beta\mtx{I}$. Further, the vector $\vec{c}=\gamma\vec{\mu}$ is scaled version of the mean. 

We can now rewrite the LLR score as
    \begin{equation}\label{eq:simpler-llr}
        \begin{aligned}
        \ell & = \vec{x}_\text{e}\transp (\alpha\mtx{I})\vec{x}_\text{t}
        + \vec{x}_\text{t}\transp (\alpha\mtx{I})\vec{x}_\text{e}
        + \vec{x}_\text{e}\transp (\beta\mtx{I})\vec{x}_\text{t}
        + \vec{x}_\text{t}\transp (\beta\mtx{I})\vec{x}_\text{e}\\
        & + \gamma\vec{\mu}\transp(\vec{x}_\text{e}+\vec{x}_\text{t}) + k \\
        & = 2\alpha\vec{x}_\text{e}\transp\vec{x}_\text{t} + \beta(\Vert \vec{x}_\text{e} \Vert^2 + \Vert \vec{x}_\text{t} \Vert^2) + \gamma\vec{\mu}\transp(\vec{x}_\text{e}+\vec{x}_\text{t}) + k.
        \end{aligned}
    \end{equation}
We proceed by noting that subtracting an arbitrary global offset $\vec{b}\in\mathbb{R}^d$ from all speaker embeddings, $\vec{x}_{i,j} \leftarrow \vec{x}_{i,j} - \vec{b}$, preserves pairwise distances and angles between vectors and therefore retains the geometric structure of the speaker embedding space. This leads to an otherwise identical PLDA model but with global mean $\vec{\mu} - \vec{b}$. Therefore, by \emph{choosing} $\vec{b}=\vec{\mu}$, the new model has a global mean $\vec{0}$ and \eqref{eq:simpler-llr} becomes
    \begin{equation}
        \ell = 2\alpha\vec{x}_\text{e}\transp\vec{x}_\text{t} + \beta(\Vert \vec{x}_\text{e} \Vert^2 + \Vert \vec{x}_\text{t} \Vert^2) + k.\nonumber
    \end{equation}
If both the enrollment and test embeddings are length-normalized ($\Vert\vec{x}_\text{e}\Vert = \Vert\vec{x}_\text{t}\Vert=1$), the score further simplies to
    \begin{equation}
        \ell = 2\alpha\vec{x}_\text{e}\transp\vec{x}_\text{t} + 2\beta + k\nonumber,
    \end{equation}
which is an affine transform of the inner product $\vec{x}_\text{e}\transp\vec{x}_\text{t}$ (cosine score) between the enrollment and test embeddings. Thus, PLDA scoring with spherical covariance matrices and cosine scoring are mappable to one another through order-preserving transform and therefore present equivalent scoring rules.

%\begin{equation}
%\begin{aligned}
%\mtx{P} & = \frac{1}{2} (\mtx{B}+\mtx{W} )^{-1}\mtx{B} (\mtx{B}+\mtx{W} - \mtx{B}(\mtx{B}+\mtx{W})^{-1}\mtx{B})^{-1}\nonumber\\
%\mtx{Q} & = \frac{1}{2} \Big[ (\mtx{B}+\mtx{W} )^{-1} - (\mtx{B}+\mtx{W} - \mtx{B}(\mtx{B}+\mtx{W})^{-1}\mtx{B})^{-1} \Big],
%\end{aligned}
%\end{equation}
%the vector 
%\begin{equation}\nonumber
%    \vec{c} =-2(\mtx{P} + \mtx{Q})\vec{\mu} ,   
%\end{equation}
%and the scalar
%\begin{equation}\nonumber
%    k = \frac{1}{2}(\log|\mtx{B} + \mtx{W}| - \log |\mtx{B} + \mtx{W} - \mtx{B}(\mtx{B} + \mtx{W})^{-1}\mtx{B}|) - \vec{c}\transp\vec{\mu}
%\end{equation}
%are functions of the PLDA parameters $(\vec{\mu}, \mtx{B}, \mtx{W})$.

%the vector $, scalar 
%\begin{equation}\label{eq:k-constant}
%, 
%\end{equation}
%and symmetric matrices $\mtx{P}$ and $\mtx{Q}$ are determined by the PLDA model parameters :
%	\begin{gather} \label{eq:Q}
%\end{equation}

\subsection{Note on PLDA and many-to-many verification} \label{appendix:plda-centroids}

Consider \emph{set-to-set} speaker verification where a trial is represented by a pair of sets with arbitrary cardinalities: $\mtx{X}_\text{e}=\{\vec{x}_{\text{e},1}, \dots, \vec{x}_{\text{e},N}\}$ and $\mtx{X}_\text{t}=\{\vec{x}_{t,1}, \dots, \vec{x}_{\text{t},M}\}$, referred to as the \emph{enrollment} and the \emph{test} set, respectively.

Assuming that the observations are generated from the PLDA model with the latent identity variable, denoted by $\vec{y}$, the verification likelihood ratio can be written as a function of the posterior
distributions of  $\vec{y}$ \cite{Villalba2017-TVAE}
\begin{equation}
    \mathrm{LR}(\mtx{X}_\text{e}, \mtx{X}_\text{t}) = \frac{p(\mtx{X}_\text{e}, \mtx{X}_\text{t}| \mathrm{same})}{p(\mtx{X}_\text{e}, \mtx{X}_\text{t}| \mathrm{diff})} \\
    = \int \frac{p(\vec{y} | \mtx{X}_\text{e}) p(\vec{y} | \mtx{X}_\text{t})}{p(\vec{y})} \dif \vec{y},
\end{equation}
where the integral admits the closed-form expression.
These posteriors can be seen to be Gaussian with the mean parameters depending only on the average of feature vectors:
\begin{equation} \nonumber
    \begin{aligned}
        p(\vec{y} | \vec{x}_{1}, \dots, \vec{x}_{N}) = \mathcal{N}(\vec{y}| \vec{m}, \mtx{\Sigma}), \\
        \mtx{\Sigma} = \left(\mtx{B}^{-1} + N \cdot \mtx{W}^{-1} \right)^{-1},  \\
        \vec{m} = N \cdot \mtx{\Sigma} \mtx{W}^{-1} \vec{c},
    \end{aligned}
\end{equation}
where $\vec{c} = \frac{1}{N} \sum_{i=1}^N \vec{x}_i$ is the centroid (average vector) of the set. One can see that the log-likelihood ratio can be computed even if both sets are represented \emph{implicitly} by pairs of (\texttt{centroid}, \texttt{cardinality}):
\begin{equation} \label{eq:plda-llr-cnt-counts}
    \mathrm{LR}(\mtx{X}_e, \mtx{X}_t) \equiv \mathrm{LR}((\vec{c}_e, N), (\vec{c}_t, M)),
\end{equation}
where $\vec{c}_e$ and $\vec{c}_t$ are the centroids of $\mtx{X}_e$ and $\mtx{X}_t$, respectively. The counts, $N$ and $M$, are known as the \emph{zero-order statistics}.

It worth noting that embeddings averaging is a commonly used strategy for speaker verification in the case of multiple enrollment utterances \cite{Rajan2014-single-to-multiple}. And PLDA can be also seen as an instance of ``centroid based'' scoring, with a difference that the information about the set cardinality is preserved. Intuitively, the information about set cardinality allows to make judgments about uncertainty: the more vectors are in the set, the higher confidence of the mean estimate. The PLDA model has advantage of taking this information into account.

Let us now recall the centroid algorithm where the speakers are represented by centroids, and the centroids are updated by the exponential smoothing rule:
\begin{equation} \nonumber
    \vec{c}^{(t)} = \alpha \vec{x}^{(t)} + (1 - \alpha)\vec{c}^{(t-1)}.
\end{equation}
By direct substitution of the update rule back into itself, one can see that it corresponds to the weighted averaging:
\begin{equation} \nonumber
    \begin{aligned}
        \vec{c}^{(N)} =\sum_{i=1}^N p_i \vec{x}^{(t)} \\
        p_i = 
        \begin{cases}
           (1 - \alpha)^{N-1} &\text{if $i = 1$} \\
           \alpha (1 - \alpha)^{(N-i)} &\text{if $i > 1$} \\
        \end{cases}
    \end{aligned}
\end{equation}
with $\vec{c}^{(0)} = \vec{x}_1$.

Now, if one wants to use the updated centroid for verification against the test utterance $\vec{x}_\text{test}$, it is not immediately obvious how to define the zero-order statistic $n$ in $\mathrm{LR}((\vec{c}^{(N)}, n), \vec{x}_\text{test})$, given by \eqref{eq:plda-llr-cnt-counts}. For example, updating a centroid with a very small $\alpha \approx 0$ does not benefit from the information available after collecting several new data points. Therefore, setting $n$ with the number of collected embeddings would not be adequate.

Here we propose a heuristic to set the count parameter in the case of unequal weights. First, let us notice that if $\alpha=0$ or $\alpha=1$, this boils down to the single enrollment verification, that is, the input count is $1$. This corresponds to either $p_1=1$ or $p_N=1$ with other weights being zero. Also, if $p_i = 1/N$ for all $i=1...N$, then the cardinality is $N$. 

Noting that the entropy of the discrete uniform distribution $\vec{p} = [p_1, p_2, ..., p_N]$ where $p_i = 1/N$ equals to $\log N$, we calculate the zero-order statistic as
\begin{equation} \nonumber
    n  = e^{\mathcal{H}(\vec{p})}.
\end{equation}
where $\mathcal{H}(\vec{p}) = - \sum_i p_i \log p_i$ is the entropy. Then, for the maximum entropy, achieved for the uniform distribution, $n = N$. At the same time, if one of the weights equals to $1$, then $n=1$, which is also consistent with the special case.

%$[\alpha, \alpha (1 - \alpha), \alpha (1 - \alpha)^2, ..., \alpha (1 - \alpha)^{N-1}, (1 - \alpha)^N]$.

Another issue related to setting the zero-order statistic is discussed in studies \cite{Stafylakis2013-correlations, McCree2017-variability, McCree2019-loo} in the context of compensating apparently inadequate assumption of feature independence. The proposed solutions suggest to scale down zero-order statistics $n_r = r\cdot n$ with the recommended factor of $r =0.2-0.3$ \cite{Stafylakis2013-correlations, Diez2020-VBx} or simply setting $n=1$ independently of the set cardinality \cite{Rajan2014-single-to-multiple}. The latter heuristic is nothing else than a commonly adopted recipe of embedding averaging.

% Q-scoring:
% ref: https://www.isca-speech.org/archive/pdfs/odyssey_2018/lopez18_odyssey.pdf
% ref: https://www.isca-speech.org/archive/pdfs/interspeech_2017/villalba17_interspeech.pdf

% Statistic scaling
% https://espace2.etsmtl.ca/id/eprint/6017/1/2013_ICASSP_CompensationForInterframe.pdf
% https://arxiv.org/pdf/2104.02469.pdf

\subsection{VB clustering details} \label{appendix:vb}

%TODO: re-write
We assume that the extracted utterance embeddings were generated by the PLDA model with the full-rank speaker subspace in the form of two-covariance model \cite{Brummer2010-two-cov}. 
This model, specified by \eqref{eq:plda-model}, is parameterized by two matrices $\mtx{B}$ and $\mtx{W}$ interpreted as between- and within-speaker covariances. We assume the data have zero-mean.

% the following probability distributions: %(see also Appendix):
% \begin{equation}
%     \begin{aligned}
%         p(\vec{y}) & = \mathcal{N}(\vec{y}|\vec{0},\mtx{B}) \\ \label{eq:2cov-likelihood}
%         p(\vec{x}|\vec{y}) & = \mathcal{N}(\vec{x}|\vec{y}, \mtx{W})
%     \end{aligned}
% \end{equation}
% where the two matrices $\mtx{B}$ and $\mtx{W}$ model between- and within-speaker covariances. 

We construct a finite mixture model for partitioning a set of embeddings $\mtx{X}=\{\vec{x}_{1}, \dots, \vec{x}_{N}\}$ into $K$ clusters corresponding to individual speakers. The speakers are represented by Gaussians with a shared covariance matrix $\mtx{W}$ and speaker-specific latent mean vectors $\mtx{Y} = \{\vec{y}_{1}, \dots, \vec{y}_{K}\}$.

Further, each embedding $\vec{x}_{i}$ is associated with a binary one-hot vector $\vec{z}_{i}$, whose components indicate whether this observation belongs to the corresponding mixture component. For instance, if the embedding $\vec{x}_{i}$ is generated from component $k$ then $z_{i,k} = 1$ with other vector components equal to zero.

The latent variables $\mtx{Z} = \{\vec{z}_{1}, \dots, \vec{z}_{N}\}$ are drawn independently from the categorical distributions governed by the mixing coefficients $\pi_k$:
\begin{equation} \nonumber
    \begin{aligned}
        p(\mtx{Z}) = \prod_{i=1}^N \prod_{k=1}^K \pi_k^{z_{i,k}}
    \end{aligned}
\end{equation}
%We assume that the prior distribution for mixture assignments is uniform: $\pi_1 = \pi_2 = ... = \pi_K$.

Conditioned on $\mtx{Z}$, the embeddings are assumed to be independently drawn from a speaker-specific Gaussian distribution so that
\begin{equation} \nonumber
    \begin{aligned}
        p(\mtx{X}|\mtx{Y}, \mtx{Z}) = \prod_{i=1}^N \prod_{k=1}^K \mathcal{N}(\vec{x}_{i}|\vec{y}_{k}, \mtx{W})^{z_{i,k}}
    \end{aligned}
\end{equation}
%where $\mtx{X}=\{\vec{x}_{1}, ..., \vec{x}_{N}\}$ is the full data set, $\mtx{Z} = \{\vec{z}_{1}, ..., \vec{z}_{N}\}$ is a set of corresponding assignments, and $\mtx{Y} = \{\vec{y}_{1}, ..., \vec{y}_{K}\}$ is a set of all speaker-specific means.

We further modified the standard mixture model described above by introducing one extra label to present `background class'. To this end, we added one component to the mixture model that represents the marginal distribution of embeddings:
\begin{equation} \nonumber
    \begin{aligned}
        p(\vec{x}) = \int p(\vec{x}|\vec{y}) p(\vec{y}) \dif \vec{y} = \mathcal{N}(\vec{x}|\vec{0}, \mtx{B} + \mtx{W})
    \end{aligned}
\end{equation}
This component, indexed as component $K+1$, corresponds to the embeddings which does not belong to any of $K$ clusters. In the context of household speaker recognition this component is intended to model all the unknown speakers (guests) at once, since discriminating those speakers is irrelevant to the task.
\begin{equation} \nonumber
    \begin{aligned}
        p(\vec{x}_{i}|\mtx{Y}, \vec{z}_{i}) = 
        \begin{cases}
          \mathcal{N}(\vec{x}_{i}|\vec{0}, \mtx{B} + \mtx{W}) &\text{if $z_{i,K+1}=1$}\\
          \mathcal{N}(\vec{x}_{i}|\vec{y}_{k}, \mtx{W})^{z_{i,k}} &\text{otherwise}\\
         \end{cases}
    \end{aligned}
\end{equation}
%% THIS is already in the main text
%To define the prior probability for the `background class', we use the threshold parameter $\tau$, which is also present in the previously describe algorithms. In detail, we define this probability as $\pi_{K+1} = \sigma(\tau)$, where $\sigma(\cdot)$ denotes the sigmoid function. Accordingly, the remaining probabilities $\pi_k$ are initialized such that $\sum_k \pi_k = 1$ for $k=1...K+1$. 

The clustering problem requires finding the most likely partition of the data, represented by $\mtx{Z}$. 
Following \cite{Diez2020-VBx}, we use variational Bayesian inference \cite{Bishop-MachineLearning2006} to approximate the posterior distribution $p(\mtx{Z}|\mtx{X}) \approx \prod_i q(\vec{z}_i)$. 
Similar to the semi-supervised k-means clustering described above, posterior distribution $q(\vec{z})$ for a subset of labeled data points are initialized by one-hot representations of their known labels and kept fixed during inference.
After the convergence, the inferred cluster assignments can be found as $k_i^* = \arg\max_k q(\vec{z_{i,k}})$. Detailed derivations and the update rules can be found in \cite{Diez2020-VBx}.

Finally, the joint distribution of all of the random variables is given by
% The complete model can be also defined in terms of the joint distribution of all of the random variables as

\begin{equation} \nonumber
    \begin{aligned}
        p(\mtx{X}, \mtx{Y}, \mtx{Z}) = p(\mtx{X}|\mtx{Y}, \mtx{Z}) p(\mtx{Y}) p(\mtx{Z}).
    \end{aligned}
\end{equation}

The clustering problem requires finding the most likely partition of the data, represented by $\mtx{Z}$. 
This can be formulated as finding the most likely partition $\mtx{Z}^{*} = \arg\max_{\mtx{Z}} p(\mtx{Z}|\mtx{X})$ where $p(\mtx{Z}|\mtx{X}) = \int  p(\mtx{Z} | \mtx{Y}, \mtx{X}) \dif \mtx{Y}$ is the clustering (partition) posterior distribution. Since direct optimization of this objective is intractable due to vast combinatorial search space, one has to resort to available approximation techniques. We use variational Bayesin inference \cite{Bishop-MachineLearning2006} to approximate the joint posterior $q(\mtx{Z}, \mtx{Y}) \approx p(\mtx{Z}, \mtx{Y}| \mtx{X})$ assuming that the approximate posterior is factorizes as $q(\mtx{Z}, \mtx{Y}) = q(\mtx{Z})q(\mtx{Y})$. We use variational inference because is often much faster than Monte-Carlo based \cite{Andrieu2003-mcmc} or greedy algorithms \cite{Silnova2020-probemb}.

Following \cite{Diez2020-VBx}, we search for such $q(\mtx{Z})$ and $q(\mtx{Y})$ that maximize the weighted \emph{evidence lower bound} (ELBO) augmented with a pair of scaling factors $F_A$ and $F_B$:
\begin{equation} \nonumber
    \begin{aligned}
        \mathcal{L}(q(\mtx{Z}), q(\mtx{Y})) = F_A\cdot \E_{q(\mtx{Z})q(\mtx{Y})} [\log p(\mtx{X}|\mtx{Y}, \mtx{Z})] \\ 
        - F_B\cdot \mathrm{KL}(q(\mtx{Y})\Vert p(\mtx{Y})) - \mathrm{KL}(q(\mtx{Z})\Vert p(\mtx{Z})).
    \end{aligned}
\end{equation}
Here, $\E_q[\cdot]$ denotes expectation with respect to $q$ and $\mathrm{KL}(\cdot\Vert\cdot)$ is the Kullback-Leibler divergence \cite{Bishop-MachineLearning2006}. Despite that the theoretically correct variational Bayesian inference assumes $F_A = F_B = 1$, tuning the values of these scaling factors helps to compensate for wrong model assumptions of statistical independence and to improve the clustering performance. For a more in-depth discussion of their interpretations, please refer to \cite{Diez2020-VBx}.

% update equations
The assumption of factorized posterior leads to the inference algorithm consisting of iterative updates of factors $q(\mtx{Z})$ and $q(\mtx{Y})$, with each update having a closed-form solution. 
Similar to the semi-supervised k-means clustering described above, posterior distribution $q(\vec{z})$ for a subset of labeled data points are initialized by one-hot representations of their known labels and kept fixed during inference.
After the convergence, the inferred cluster assignments can be found as $k_i^* = \arg\max_k q(\vec{z_{i,k}})$. Detailed derivations and the update rules can be found in \cite{Diez2020-VBx}.

%\fi

% https://tex.stackexchange.com/questions/131087/displaying-authors-name-in-a-bibliographic-entry-in-the-form-surname-first-in
% https://tex.stackexchange.com/questions/160862/bibtex-authors-surname-followed-by-multiple-initials
% https://tex.stackexchange.com/questions/47600/modify-list-of-authors-format-names-problem
% https://tex.stackexchange.com/questions/127705/natbib-how-to-display-partial-authors-in-reference

% \bibliographystyle{IEEEbib}
% \bibliography{Odyssey2022_BibEntries}
\section{References}
{
% \setstretch{0.88}
\printbibliography

@book{esl_textbook,
  address = {New York, NY, USA},
  author = {Hastie, Trevor and Tibshirani, Robert and Friedman, Jerome},
  publisher = {Springer New York Inc.},
  series = {Springer Series in Statistics},
  title = {The Elements of Statistical Learning},
  year = 2001
}

@inproceedings{Leeuwen2014-calibration,
  author    = {David A. van Leeuwen and
               Niko Brummer and
               Albert Swart},
  title     = {A comparison of linear and non-linear calibrations for speaker recognition},
  booktitle = {Odyssey},
  year      = {2014}
}

@inproceedings{openset_sid2006,
author={A. M. Ariyaeeinia, J. Fortuna, P. Sivakumaran and A. Malagaonkar},
title={Verification effectiveness in open-set speaker identification},
booktitle={{ICVISP}},
year={2006},
volume={153},
number={5},
pages={618-624}
}

@inproceedings{tslearning2019,
  author={Wang, Shuai and Yang, Yexin and Wang, Tianzhe and Qian, Yanmin and Yu, Kai},
  booktitle={{ICASSP}}, 
  title={Knowledge Distillation for Small Foot-print Deep Speaker Embedding}, 
  year={2019},
  pages={6021--6025}
}

@inproceedings{jer_original2019,
  author={Neville Ryant and Kenneth Church and Christopher Cieri and Alejandrina Cristia and Jun Du and Sriram Ganapathy and Mark Liberman},
  title={{The Second DIHARD Diarization Challenge: Dataset, Task, and Baselines}},
  year={2019},
  booktitle={Interspeech},
  pages={978--982}
}

@inproceedings{Zhou2003-local-global,
  author    = {Dengyong Zhou and
               Olivier Bousquet and
               Thomas Navin Lal and
               Jason Weston and
               Bernhard Sch{\"{o}}lkopf},
  title     = {Learning with Local and Global Consistency},
  booktitle = {{NIPS}},
  pages     = {321--328},
  year      = {2003}
}

@article{Chen2021-Label-Propagation,
  author    = {Long Chen and
               Venkatesh Ravichandran and
               Andreas Stolcke},
  title     = {Graph-based Label Propagation for Semi-Supervised Speaker Identification},
  journal   = {CoRR},
  volume    = {abs/2106.08207},
  year      = {2021},
  url       = {https://arxiv.org/abs/2106.08207}
}

@inproceedings{resnet34,
  author={He, Kaiming and Zhang, Xiangyu and Ren, Shaoqing and Sun, Jian},
  booktitle={{CVPR}}, 
  title={Deep Residual Learning for Image Recognition}, 
  year={2016},
  pages={770-778}
}

@article{clova,
  title={Clova baseline system for the {VoxCeleb} Speaker Recognition Challenge 2020},
  author={Heo, Hee Soo and Lee, Bong-Jin and Huh, Jaesung and Chung, Joon Son},
  journal={arXiv preprint arXiv:2009.14153},
  year={2020}
}

@misc{speechbrain,
      title={{SpeechBrain}: A General-Purpose Speech Toolkit}, 
      author={Mirco Ravanelli and Titouan Parcollet and Peter Plantinga and Aku Rouhe and Samuele Cornell and Loren Lugosch and Cem Subakan and Nauman Dawalatabad and Abdelwahab Heba and Jianyuan Zhong and Ju-Chieh Chou and Sung-Lin Yeh and Szu-Wei Fu and Chien-Feng Liao and Elena Rastorgueva and Francois Grondin and William Aris and Hwidong Na and Yan Gao and Renato De Mori and Yoshua Bengio},
      year={2021},
      eprint={2106.04624},
      archivePrefix={arXiv},
      primaryClass={eess.AS}
}

@inproceedings{Tan2021-shared-device-subset,
  author    = {Zhenning Tan and
               Yuguang Yang and
               Eunjung Han and
               Andreas Stolcke},
  title     = {Improving Speaker Identification for Shared Devices by Adapting Embeddings
               to Speaker Subsets},
  booktitle = {{ASRU}},
  pages     = {1124--1131},
  year      = {2021}
}

@inproceedings{ts_learning,
  author={Ng, Raymond W. M. and Liu, Xuechen and Swietojanski, Pawel},
  booktitle={{IEEE} {SLT} Workshop}, 
  title={Teacher-Student Training for Text-Independent Speaker Recognition}, 
  year={2018},
  pages={1044-1051}
}

@article{WANG2020101114,
title = {{ASVspoof} 2019: A large-scale public database of synthesized, converted and replayed speech},
journal = {Computer Speech and Language},
volume = {64},
pages = {101114},
year = {2020},
url = {https://www.sciencedirect.com/science/article/pii/S0885230820300474},
author = {Xin Wang and Junichi Yamagishi and Massimiliano Todisco and Hector Delgado and Andreas Nautsch and Nicholas Evans and Md Sahidullah and Ville Vestman and Tomi Kinnunen and Kong Aik Lee and Lauri Juvela and Paavo Alku and Yu-Huai Peng and Hsin-Te Hwang and Yu Tsao and Hsin-Min Wang and Sebastien Le Maguer and Markus Becker and Fergus Henderson and Rob Clark and Yu Zhang and Quan Wang and Ye Jia and Kai Onuma and Koji Mushika and Takashi Kaneda and Yuan Jiang and Li-Juan Liu and Yi-Chiao Wu and Wen-Chin Huang and Tomoki Toda and Kou Tanaka and Hirokazu Kameoka and Ingmar Steiner and Driss Matrouf and Jean-Francois Bonastre and Avashna Govender and Srikanth Ronanki and Jing-Xuan Zhang and Zhen-Hua Ling}
}

@article{Rajan2014-single-to-multiple,
title = {From single to multiple enrollment i-vectors: Practical {PLDA} scoring variants for speaker verification},
journal = {Digital Signal Processing},
volume = {31},
pages = {93-101},
year = {2014},
issn = {1051-2004},
doi = {https://doi.org/10.1016/j.dsp.2014.05.001},
url = {https://www.sciencedirect.com/science/article/pii/S1051200414001377},
author = {Padmanabhan Rajan and Anton Afanasyev and Ville Hautamaki and Tomi Kinnunen}
}

@inproceedings{Brummer2010-two-cov,
  author    = {Niko Br{\"{u}}mmer and
               Edward de Villiers},
  title     = {The speaker partitioning problem},
  booktitle = {Odyssey},
  pages     = {34},
  year      = {2010}
}

@inproceedings{Burget2011-discriminatively-trained-PLDA,
  author    = {Luk{\'{a}}s Burget and
               Oldrich Plchot and
               Sandro Cumani and
               Ondrej Glembek and
               Pavel Matejka and
               Niko Br{\"{u}}mmer},
  title     = {Discriminatively trained Probabilistic Linear Discriminant Analysis
               for speaker verification},
  booktitle = {{ICASSP}},
  pages     = {4832--4835},
  year      = {2011}
}

@inproceedings{Corduneanu2001-mixture,
author = {Corduneanu, A. and Bishop, Christopher},
title = {Variational Bayesian Model Selection for Mixture Distributions},
booktitle = {{AISTATS}},
year = {2001},
url = {https://www.microsoft.com/en-us/research/publication/variational-bayesian-model-selection-for-mixture-distributions/},
pages = {27--34}
}

@inproceedings{Rohdin-2014,
  author    = {Rohdin, Johan and
               Biswas, Sangeeta and
               Shinoda, Koichi},
  title     = {Discriminative {PLDA} training with application-specific loss functions for speaker verification},
  booktitle = {Odyssey}, 
  pages     = {26--32},
  year      = {2014},
}

@inproceedings{voxceleb1,
  author={A. Nagrani and J. Chung and A. Zisserman},
  title={{VoxCeleb}: A Large-Scale Speaker Identification Dataset},
  year={2017},
  booktitle={Interspeech},
  pages={2616--2620}
}

@inproceedings{Sizov2014-unifying,
  author    = {Aleksandr Sizov and
               Kong{-}Aik Lee and
               Tomi Kinnunen},
  title     = {Unifying Probabilistic Linear Discriminant Analysis Variants in Biometric
               Authentication},
  booktitle = {{S+SSPR}, Joint {IAPR} International Workshop},
  series    = {Lecture Notes in Computer Science},
  volume    = {8621},
  pages     = {464--475},
  publisher = {Springer},
  year      = {2014}
}

@inproceedings{voxceleb2,
  author={J. Chung and A. Nagrani and A. Zisserman},
  title={{VoxCeleb2}: deep Speaker Recognition},
  year=2018,
  booktitle={Interspeech},
  pages={1086--1090}
}

@inproceedings{ecapa-tdnn,
  author    = {Brecht Desplanques and
               Jenthe Thienpondt and
               Kris Demuynck},
  title     = {{ECAPA-TDNN:} Emphasized Channel Attention, Propagation and Aggregation
               in {TDNN} Based Speaker Verification},
  booktitle = {Interspeech},
  pages     = {3830--3834},
  year      = {2020},
}

@inproceedings{astats_pooling,
  author={K. Okabe and T. Koshinaka and K. Shinoda},
  title={Attentive Statistics Pooling for Deep Speaker Embedding},
  year={2018},
  booktitle={Interspeech},
  pages={2252--2256}
}

@ARTICLE{aam_softmax,
  author={Wang, F. and Cheng, J. and Liu, W. and Liu, H.},
  journal={IEEE Signal Processing Letters}, 
  title={Additive Margin Softmax for Face Verification}, 
  year={2018},
  volume={25},
  number={7},
  pages={926-930},
  doi={10.1109/LSP.2018.2822810}}

@inproceedings{Snyder_etdnn_2019,
  author={D. {Snyder} and D. {Garcia-Romero} and G. {Sell} and A. {McCree} and D. {Povey} and S. {Khudanpur}},
  booktitle={{ICASSP}}, 
  title={Speaker Recognition for Multi-speaker Conversations Using {X}-vectors}, 
  year={2019},
  pages={5796--5800}
}

@article{GREENBERG2020-two-decades,
title = {Two decades of speaker recognition evaluation at the national institute of standards and technology},
journal = {Computer Speech and Language},
volume = {60},
pages = {101032},
year = {2020},
issn = {0885-2308},
doi = {https://doi.org/10.1016/j.csl.2019.101032},
url = {https://www.sciencedirect.com/science/article/pii/S0885230819302761},
author = {Craig S. Greenberg and Lisa P. Mason and Seyed Omid Sadjadi and Douglas A. Reynolds}
}

@article{Speaker-rec-deep-learning-review-2021,
title = {Speaker recognition based on deep learning: An overview},
journal = {Neural Networks},
volume = {140},
pages = {65-99},
year = {2021},
author = {Z. Bai and X. Zhang}
}

@article{Zeng2021-attmultienroll,
  author    = {Chang Zeng and
               Xin Wang and
               Erica Cooper and
               Junichi Yamagishi},
  title     = {Attention Back-end for Automatic Speaker Verification with Multiple
               Enrollment Utterances},
  journal   = {CoRR},
  volume    = {abs/2104.01541},
  year      = {2021},
  url       = {https://arxiv.org/abs/2104.01541}
}

@inproceedings{coralplus, 
author={Lee, Kong Aik and Wang, Qiongqiong and Koshinaka, Takafumi},  
booktitle={{ICASSP}},   
title={The {CORAL+} Algorithm for Unsupervised Domain Adaptation of {PLDA}},   
year={2019},  
pages={5821--5825}
}

@inproceedings{coral,
author = {Sun, Baochen and Feng, Jiashi and Saenko, Kate},
title = {Return of Frustratingly Easy Domain Adaptation},
year = {2016},
booktitle = {{AAAI}},
pages = {2058--2065}
}

@inproceedings{Aronowitz2014-idvc,
  author    = {Hagai Aronowitz},
  title     = {Compensating Inter-Dataset Variability in {PLDA} Hyper-Parameters
               for Robust Speaker Recognition},
  booktitle = {Odyssey},
  year      = {2014}
}

@inproceedings{kaldi_aplda,
  author={Pierre-Michel Bousquet and Mickael Rouvier},
  title={{On Robustness of Unsupervised Domain Adaptation for Speaker Recognition}},
  year={2019},
  booktitle={Interspeech},
  pages={2958--2962}
}

@inproceedings{coral_asv2018,
  author={Md Jahangir Alam and Gautam Bhattacharya and Patrick Kenny},
  title={{Speaker Verification in Mismatched Conditions with Frustratingly Easy Domain Adaptation	}},
  year=2018,
  booktitle={Odyssey},
  pages={176--180}
}

@inproceedings{voices,
  author={Mahesh Kumar Nandwana and Michael Lomnitz and Colleen Richey and Mitchell McLaren and Diego Castan and Luciana Ferrer and Aaron Lawson},
  title={{The VOiCES from a Distance Challenge 2019: Analysis of Speaker Verification Results and Remaining Challenges}},
  year=2020,
  booktitle={Odyssey},
  pages={165--170}
}

@article{voxsrc-2019,
  author    = {Joon Son Chung and
               Arsha Nagrani and
               Ernesto Coto and
               Weidi Xie and
               Mitchell McLaren and
               Douglas A. Reynolds and
               Andrew Zisserman},
  title     = {{VoxSRC} 2019: The first {VoxCeleb} Speaker Recognition Challenge},
  journal   = {CoRR},
  volume    = {abs/1912.02522},
  year      = {2019}
}

@article{voxsrc-2020,
  author    = {Arsha Nagrani and
               Joon Son Chung and
               Jaesung Huh and
               Andrew Brown and
               Ernesto Coto and
               Weidi Xie and
               Mitchell McLaren and
               Douglas A. Reynolds and
               Andrew Zisserman},
  title     = {{VoxSRC} 2020: The Second {VoxCeleb} Speaker Recognition Challenge},
  journal   = {CoRR},
  volume    = {abs/2012.06867},
  year      = {2020}
}

@inproceedings{deng2018arcface,
title={{ArcFace}: Additive Angular Margin Loss for Deep Face Recognition},
author={Deng, Jiankang and Guo, Jia and Niannan, Xue and Zafeiriou, Stefanos},
booktitle={{CVPR}},
year={2019}
}

@inproceedings{chung2020in,
  title={In defence of metric learning for speaker recognition},
  author={Chung, Joon Son and Huh, Jaesung and Mun, Seongkyu and Lee, Minjae and Heo, Hee Soo and Choe, Soyeon and Ham, Chiheon and Jung, Sunghwan and Lee, Bong-Jin and Han, Icksang},
  booktitle={Interspeech},
  pages={2977--2981},
  year={2020}
}

@inproceedings{Villalba2017-TVAE,
  author    = {Jes{\'{u}}s Villalba and
               Niko Br{\"{u}}mmer and
               Najim Dehak},
  title     = {Tied Variational Autoencoder Backends for i-Vector Speaker Recognition},
  booktitle = {Interspeech},
  pages     = {1004--1008},
  year      = {2017}
}

@inproceedings{Silnova2020-probemb,
  author    = {Anna Silnova and
               Niko Brummer and
               Johan Rohdin and
               Themos Stafylakis and
               Luk{\'{a}}s Burget},
  title     = {Probabilistic Embeddings for Speaker Diarization},
  booktitle = {Odyssey},
  pages     = {24--31},
  year      = {2020}
}

@article{Andrieu2003-mcmc,
  author    = {Christophe Andrieu and
               Nando de Freitas and
               Arnaud Doucet and
               Michael I. Jordan},
  title     = {An Introduction to {MCMC} for Machine Learning},
  journal   = {Mach. Learn.},
  volume    = {50},
  number    = {1-2},
  pages     = {5--43},
  year      = {2003},
  url       = {https://doi.org/10.1023/A:1020281327116},
  doi       = {10.1023/A:1020281327116},
  bibsource = {dblp computer science bibliography, https://dblp.org}
}

@BOOK{Bishop-MachineLearning2006,
    AUTHOR={C.M. Bishop},
    TITLE={Pattern Recognition and Machine Learning},
    PUBLISHER={Springer Science+Business Media, LLC},
    ADDRESS={New York},
    YEAR={2006},
}

@BOOK{Brown1963-smoothing,
    AUTHOR={Brown, Robert Goodell},
    TITLE={Smoothing, Forecasting and Prediction of Discrete Time Series},
    PUBLISHER={Prentice-Hall},
    ADDRESS={Englewood-Cliffs, NJ},
    YEAR={1963},
}

@techreport{Zhu2005-ssl,
  author = {Zhu, Xiaojin},
  institution = {Computer Sciences, University of Wisconsin-Madison},
  number = 1530,
  title = {Semi-Supervised Learning Literature Survey},
  url = {http://pages.cs.wisc.edu/~jerryzhu/pub/ssl_survey.pdf},
  year = 2005
}

@BOOK{AKJain1999-clustering,
  address = {New York, NY, USA},
  author = {Jain, A. K. and Murty, M. N. and Flynn, P. J.},
  description = {Data clustering},
  journal = {ACM Comput. Surv.},
  number = 3,
  pages = {264--323},
  publisher = {ACM},
  title = {Data clustering: a review},
  volume = 31,
  year = 1999
}

@article{Diez2020-VBx,
  author    = {Mireia D{\'{i}}ez and
               Luk{\'{a}}s Burget and
               Federico Landini and
               Jan Cernock{\'{y}}},
  title     = {Analysis of Speaker Diarization Based on Bayesian {HMM} With Eigenvoice
               Priors},
  journal   = {{IEEE} {ACM} Trans. Audio Speech Lang. Process.},
  volume    = {28},
  pages     = {355--368},
  year      = {2020}
}

@inproceedings{Zhou2003-labelprop,
  author    = {Dengyong Zhou and
               Olivier Bousquet and
               Thomas Navin Lal and
               Jason Weston and
               Bernhard Sch{\"{o}}lkopf},
  title     = {Learning with Local and Global Consistency},
  booktitle = {{NIPS}},
  pages     = {321--328},
  year      = {2003}
}

@inproceedings{McCree2019-loo,
  author    = {Alan McCree and
               Gregory Sell and
               Daniel Garcia{-}Romero},
  title     = {Speaker Diarization Using Leave-One-Out Gaussian {PLDA} Clustering
               of {DNN} Embeddings},
  booktitle = {Interspeech},
  pages     = {381--385},
  year      = {2019}
}

@inproceedings{McCree2017-variability,
  author    = {Alan McCree and
               Gregory Sell and
               Daniel Garcia{-}Romero},
  title     = {Extended Variability Modeling and Unsupervised Adaptation for {PLDA}
               Speaker Recognition},
  booktitle = {Interspeech},
  pages     = {1552--1556},
  year      = {2017}
}

@inproceedings{Stafylakis2013-correlations,
  author    = {Themos Stafylakis and
               Patrick Kenny and
               Vishwa Gupta and
               Pierre Dumouchel},
  title     = {Compensation for inter-frame correlations in speaker diarization and
               recognition},
  booktitle = {{ICASSP}},
  pages     = {7731--7735},
  year      = {2013}
}

@inproceedings{Snyder2018-xvec,
  author    = {David Snyder and
               Daniel Garcia{-}Romero and
               Gregory Sell and
               Daniel Povey and
               Sanjeev Khudanpur},
  title     = {{X}-Vectors: Robust {DNN} Embeddings for Speaker Recognition},
  booktitle = {{ICASSP}},
  pages     = {5329--5333},
  year      = {2018}
}

@article{BUT-voxsrc2019,
  author    = {Hossein Zeinali and
               Shuai Wang and
               Anna Silnova and
               Pavel Matejka and
               Oldrich Plchot},
  title     = {{BUT} System Description to {VoxCeleb} Speaker Recognition Challenge
               2019},
  journal   = {CoRR},
  volume    = {abs/1910.12592},
  year      = {2019},
  url       = {http://arxiv.org/abs/1910.12592}
}
}

% This could be also done as follows:
%
%\begin{thebibliography}{10}
%\bibitem[1]{aluisio2001learn}Sandra M. Alu\'{i}sio, Iris Barcelos, Jandir Sampaio, and Osvaldo
%N. Oliveira Jr, ``How to learn the many unwritten
%``rules of the game'' of the academic discourse: a hybrid
%approach based on critiques and cases to support scientific
%writing,'' in Proceedings of the IEEE International Conference
%on Advanced Learning Technologies, Madison, USA,
%August 2001, pp. 257-260.
%\bibitem[2]{swales1987writing} John Swales and Hazem Najjar, ``The writing of research
%article introductions,'' Written communication, vol. 4, no.
%2, pp. 175-191, 1987.
%\bibitem[3]{day2012write} Robert Day and Barbara Pastel, How to write and publish
%a scientific paper, Cambridge University Press, 2012.
%\bibitem[4]{teufel2000} Simone Teufel, Argumentative zoning: information extraction
%from scientific text, Ph.D. thesis, University of Edinburgh,
%2000.
%\bibitem[5]{berkenkotter1989social} Carol Berkenkotter, Thomas N. Huckin, and John Ackerman,
%``Social context and socially constructed texts: The
%initiation of a graduate student into a writing research community.
%technical report no. 33.,'' Tech. Rep., Center for
%the Study of Writing, University of California Berkeley \&
%Carnegie Mellon University, 1989.
%\end{thebibliography}

\end{document}